# Title: The 2024 Motile Active Matter Roadmap

**Authors**

Gerhard Gompper[1], Howard A. Stone[2], Christina Kurzthaler[3], David Saintillan[4], Fernado Peruani[5], Dmitry A. Fedosov[1], Thorsten Auth[1], Cecile Cottin-Bizonne[6], Christophe Ybert[6], Eric Clément[7,8], Thierry Darnige[7], Anke Lindner[7], Raymond E. Goldstein[9], Benno Liebchen[10], Jack Binysh[11], Anton Souslov[12], Lucio Isa[13], Roberto di Leonardo[14], Giacomo Frangipane[14], Hongri Gu[15], Bradley J. Nelson[16], Fridtjof Brauns[17], M. Cristina Marchetti[18], Frank Cichos[19], Veit-Lorenz Heuthe[15], Clemens Bechinger[15], Amos Korman[20], Ofer Feinerman[21], Andrea Cavagna[22,23], Irene Giardina[23,22], Hannah Jeckel[24], Knut Drescher[25]

[1] Theoretical Physics of Living Matter, Institute for Advanced Simulation, Forschungszentrum Jülich, Jülich, Germany
[2] Princeton University, Department of Mechanical and Aerospace Engineering, Princeton, New Jersey, USA
[3] Max Planck Institute for the Physics of Complex Systems; Center for Systems Biology Dresden; Cluster of Excellence, Physics of Life, TU Dresden; Dresden, Germany
[4] University of California San Diego, USA
[5] CY Cergy Paris University, France
[6] Université de Lyon, Université Claude Bernard Lyon 1, CNRS, Institut Lumière Matière, Villeurbanne, France
[7] Laboratoire PMMH-ESPCI, UMR 7636 CNRS-PSL-Research University, Sorbonne Université, Université Paris Cité, Paris, France.
[8] Institut Universitaire de France, Paris, France
[9] Department of Applied Mathematics and Theoretical Physics, University of Cambridge, Cambridge, UK
[10] Technische Universität Darmstadt, 64289 Darmstadt, Germany
[11] Institute of Physics, Universiteit van Amsterdam, Science Park 904, 1098 XH Amsterdam, Netherlands
[12] T.C.M. Group, Cavendish Laboratory, University of Cambridge, Cambridge CB3 0HE, United Kingdom
[13] Laboratory for Soft Materials and Interfaces, Department of Materials, ETH Zurich, 8093 Zurich, Switzerland
[14] Dipartimento di Fisica, Sapienza Università di Roma, Italy
[15] Department of Physics, University of Konstanz, Konstanz, Germany
[16] Institute of Robotics and Intelligent Systems, ETH Zürich, Zurich, Switzerland
[17] Kavli Institute for Theoretical Physics, University of California Santa Barbara, California 93106, USA
[18] Department of Physics, University of California Santa Barbara, California 93106, USA
[19] Molecular Nanophotonics, Leipzig University, 04013 Leipzig Germany
[20] University of Haifa, Department of Computer Science, Haifa, Israel, and French National Center for Scientific Research (CNRS), UMI FILOFOCS, Israel
[21] Department of Physics of Complex Systems, Weizmann Institute of Science, Israel
[22] Istituto Sistemi Complessi (ISC-CNR), Rome, Italy
[23] Dipartimento di Fisica, Sapienza Università di Roma & INFN, Unità di Roma 1, Rome, Italy
[24] Department of Biology and Biological Engineering, California Institute of Technology, 91125 Pasadena, USA
[25] Biozentrum, University of Basel, 4056 Basel, Switzerland

Correspondence: g.gompper@fz-juelich.de

**Abstract:**

Activity and autonomous motion are fundamental aspects of many living and engineering systems. Here, the scale of biological agents covers a wide range, from nanomotors, cytoskeleton, and cells, to insects, fish, birds, and people. Inspired by biological active systems, various types of autonomous synthetic nano- and micromachines have been designed, which provide the basis for multifunctional, highly responsive, intelligent active materials. A major challenge for understanding and designing active matter is their inherent non-equilibrium nature due to persistent energy consumption, which invalidates equilibrium concepts such as free energy, detailed balance, and time-reversal symmetry. Furthermore, interactions in ensembles of active agents are often non-additive and non-reciprocal. An important aspect of biological agents is their ability to sense the environment, process this information, and adjust their motion accordingly. It is an important goal for the engineering of micro-robotic systems to achieve similar functionality.

With many fundamental properties of motile active matter now reasonably well understood and under control, the ground is prepared for the study of physical aspects and mechanisms of motion in complex environments, of the behavior of systems with new physical features like chirality, of the development of novel micromachines and microbots, of the emergent collective behavior and swarming of intelligent self-propelled particles, and of particular features of microbial systems.

The vast complexity of phenomena and mechanisms involved in the self-organization and dynamics of motile active matter poses major challenges, which can only be addressed by a truly interdisciplinary effort involving scientists from biology, chemistry, ecology, engineering, mathematics, and physics. The 2024 motile active matter roadmap of Journal of Physics: Condensed Matter reviews the current state of the art of the field and provides guidance for further progress in this fascinating research area.

## Contents

# Introduction

Gerhard Gompper, Theoretical Physics of Living Matter, Institute for Advanced Simulation, Forschungszentrum Jülich, D-52425 Jülich, Germany

Motile active matter is a novel class of nonequilibrium systems composed of a large number of autonomous self-propelled agents. The scale of agents ranges from nanomotors, microswimmers and cells, to ants, fish, birds, and humans. Unraveling, predicting, and controlling the behavior of active matter is a truly interdisciplinary endeavor at the interface of biology, chemistry, ecology, engineering, mathematics, and physics [1]. Since the pioneering work of Gray and Hancock [2], Berg [3], and Purcell [4], the investigation and understanding of these systems has made enormous progress [5-10]. Methodological and theoretical progress has facilitated new insights into the fabrication of synthetic nano- and microswimmers, the propulsion mechanisms of biological and synthetic active agents, and their collective behavior [1,5-10]. With many fundamental properties of motile active matter now reasonably well understood and under control, the ground is prepared for the design of more advanced machines with new physical features, the development of novel synthetic active materials, the study of the behavior in complex environments, and the dynamics of swarms of agents with directional sensing and non-reciprocal interactions. This roadmap provides an overview of the current state-of-the-art, and provides a perspective of future research directions on natural and artificial active agents and their collective properties.

**Active Matter in Complex Environments**
Biological microorganisms are exposed to a large variety of environments, from the soil and the ocean to blood and tissues. Synthetic microswimmers, which self-propel in microstructured aqueous environments, have myriad potential applications, e.g., in cell sorting and manipulation, micro-manufacturing, and assembly of dynamic and programmable materials. Complex environments encompass geometric confinement, fluid-host media, or the combination of both. Understanding the individual and collective behavior of active matter in realistic inhomogeneous environments is therefore absolutely essential for all real-world applications.

**Chiral Active Matter**
Chirality is an intrinsic fundamental property of many natural and artificial systems. Understanding the role of chirality in the active, non-equilibrium dynamics of interacting many-body systems is a major challenge. Artificial microsystems driven out-of-equilibrium by external torques are ideal model systems to investigate these phenomena, since they avoid the inherent complexity of biological active matter. Spinning particles dispersed in a fluid represent a special class of artificial active systems that inject vorticity at the microscopic level.

**Micromachines and Microbots**
Intelligent micromachines, with dimensions ranging from a few millimeters down to hundreds of nanometers, are miniature systems capable of performing specific tasks autonomously at small scales. A longstanding challenge in the design of microbots is to endow them with a semblance of "intelligence": the ability to autonomously sense the local microenvironment, e.g., their position with respect to confining surfaces, and respond according to their design. This promises applications in minimally invasive medicine, bioengineering, water cleaning, analytical chemistry, and more.

**Emergent collective behavior and swarming of intelligent self-propelled particles**
Group formation and swarming in living systems typically results from a delicate balance of repulsive, aligning, and attractive interactions. Here, directional sensing (e.g. visual perception), non-reciprocal interactions, and self-steering in response to gathered information leads to motility-induced cohesion and group formation. These mechanisms can be relevant not only for the self-organization of living systems, but also for the design of robust and scalable autonomous systems.

**Physics of Microbial Motility**

Cell swimming underpins a wide range of fundamental biological phenomena from microbial grazing at the base of the food web, to parasitic infections, and animal reproduction. Advances due to new experimental, theoretical, and numerical tools provide fundamental new insights, from the constraints on single-cell propulsion to the optimality of responses to environmental clues, and promise new technologies based on the control of microbial movement.

# 1 -- Motile objects in complex environments: Transport and feedback

Howard A. Stone[1] and Christina Kurzthaler[2]
[1]Princeton University, Department of Mechanical and Aerospace Engineering, Princeton, New Jersey 08544, USA
[2]Max Planck Institute for the Physics of Complex Systems; Center for Systems Biology Dresden, and Cluster of Excellence, Physics of Life, TU Dresden; Dresden, Germany

**Status**
Nature provides numerous exemplars of motile matter across various scales -- from wildebeest herds, flocks of birds, schools of fish, rafts of ants, and entangled worm blobs at the macroscale, to swimming microorganisms, suspensions of gliding biofilament/motor proteins, and synchronously beating cilia carpets at the micro/nano scales -- which have sparked novel ideas for the design of synthetic (micro)robots [1]. The transport behaviors of these active agents are dictated by their surrounding environments: fish are subject to turbulent flows impacting their coherent motions, sperm efficiently navigate the confined female reproductive tract, microorganisms spread in porous soil [3,4] (Fig. 1) and squeeze through the narrow channels of teeth and tissues, worms rapidly disentangle in response to environmental stresses [2] (Fig. 2**A**), and cytoskeletal filaments arrange inside cells to achieve diverse biological functions [5]. These manifold topics emerge at the interface of biology, agriculture, health, reproduction, respiration, metabolism, and engineering, rendering the study of motile objects in complex environments particularly relevant.

Over decades much research has focused on the physics and fluid mechanics of free swimming, yet the study of agents close to boundaries and the influence of complex media, such as porous structures and non-Newtonian rheology, on large-scale motion has been made possible only recently due to advances in technological, computational, theoretical, and data-analysis techniques. To draw connections to biological systems, it is important to characterize how quantities, such as the swimming speed, re-orientations, large-scale effective diffusion, and aggregation, are influenced by the geometric features of a porous structure, the rheological characteristics of the liquid phase, which is dictated by the molecular weight and structure of polymers in solution, and chemical fields that are self-generated via, e.g., quorum sensing, or externally imposed.

Recent observations, ranging from enhanced transport of active filaments due to entanglement [6] (Fig. 2**B**) to the suppression of motility-induced phase separation (MIPS) due to chemotaxis [7] (Fig. 2**C**) to the production of corner flows due to surfactant-producing bacteria [4] (Fig. 1**B**), demonstrate the rich physics arising in these complex non-equilibrium systems. The presence of multiple time and length scales inherent to the environment (e.g., the pore scale dimensions, the relaxation time of flexible polymers, and the diffusive time of chemicals) and the motile agents (e.g., the object size, the run length, re-orientation times, and production time of signalling molecules) makes these systems fascinating but challenging and therefore requires future development of theoretical and experimental methods.

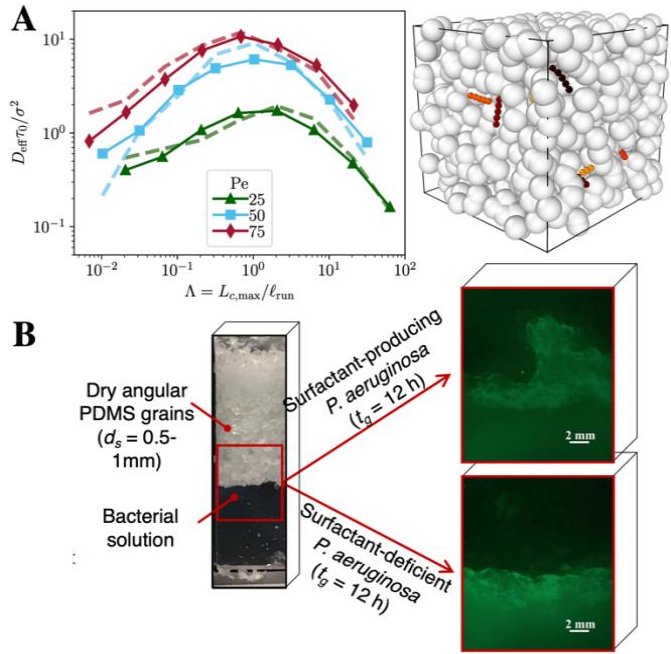

*Figure 1 **Transport in porous media**. **A**. Geometric criterion for optimal transport of active agents in porous media. Effective diffusivities $D_{eff}$ obtained from simulations of self-propelled polymers in 3D porous media (see right panel for a simulation snapshot) and compared to a theory. Reproduced and adapted from Ref. [3] under the Creative Commons Attribution License. **B**. Corner flow in unsaturated porous media generated by surfactant-producing bacteria, which communicate via quorum sensing and where the surfactant changes wettability of the surface to drive spreading. Reproduced and adapted from Ref. [4]. Copyright (2021) National Academy of Sciences.*

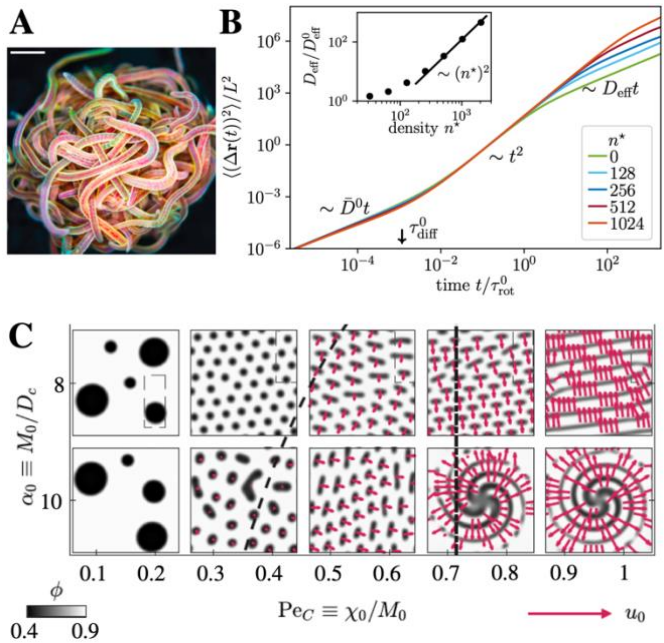

*Figure 2 **From entanglement to chemotaxis**. **A**. Entangled blob of California blackworms dissolves rapidly under environmental stress (scale bar is 3 mm). From Ref. [2]. Reprinted with permission from AAAS. **B**. Crowding-enhanced diffusion of self-propelled filaments in a highly-entangled environment of other active agents. Mean-square displacements $\langle(\Delta r(t))^2\rangle$ for different reduced number densities $n^*$ as a function of time $t$. Reprinted with permission from Ref. [6]. Copyright (2020) by the American Physical Society. **C**. Suppression of MIPS due to chemotaxis. Here, $\alpha_0 = M_0/D_c$ measures the relative importance of the active diffusivity $M_0$ and the chemical diffusivity $D_c$ and $Pe_C = \chi_0/M_0$ is the ratio of the chemotactic coefficient $\chi_0$ and $M_0$. Reprinted and adapted with permission from Ref. [7]. Copyright (2023) by the American Physical Society and courtesy of S.S. Datta.*

**Current and Future Challenges**

There are distinct forward-looking research challenges for active matter in complex environments relevant to health, medicine, and agriculture as well as the design and application of soft robotics. We believe there is value both in recognizing the conceptual theoretical and simulation framework relevant for understanding individual behavior as well as collective responses. We highlight:

1. *Towards predicting universal behaviors*. The dynamics of swimming microorganisms and synthetic active colloids often reveal unanticipated phenomena, such as enhancement of swim speed in non-Newtonian fluids, hopping-and-trapping motility in porous media (Fig.1**A**), and MIPS [1,8]. Besides the recognition that some behaviors, such as swimming and re-orientation mechanisms, appear across all domains of microbial life, we believe two current challenges are to achieve an overarching quantitative characterization of these active systems in various environmental conditions to identify universal and system-specific behaviors and to provide easy-to-use, open-source computational tools for the community to explore the wide variety of systems of interest.

2. *Bridging multiple length and time scales*. From intracellular mechanisms, such as chemotactic signalling pathways or the force-generation via lamellipodia, and the fluid dynamics of beating flagella to swimming/crawling cells, entails non-equilibrium processes at several scales. Understanding the feedback between environmental cues (chemical/mechanical) and transport behaviors represents an important step towards the design of synthetic cells and microrobots with, e.g., surface sensing mechanisms.

3. *Complex surfaces.* Most studies are restricted to dynamics near planar surfaces while microbial habitats are rough, three-dimensional, and contain complex (deformable) curved surfaces, which need to be accounted for in the context of surface accumulation.

4. *Physicochemical effects.* Microorganisms and eukaryotic cells exploit biophysical mechanisms for foraging and escaping from harm via temporal or spatial sensing of chemical gradients [9]. While such responses have been mainly studied at the single-cell level over short times, future research should investigate how spatial heterogeneity of chemical fields, produced by confinement and/or hydrodynamic flows, leads to the formation of bacterial communities in ecological niches. Further, chemical fields can impact transport of colloidal objects, including bacteria, via *diffusiophoresis* and, hence, deciphering both contributions is an important challenge.

5. *Heterogeneous suspensions*. Future work should address heterogeneous suspensions, such as diffusing colloids and active agents or active and passive biopolymers, whose motions couple via hydrodynamics, steric effects, and/or chemical fields. The latter can entail non-reciprocal interactions or entangled collective lifeforms, leading to new physics and material properties. For example, unraveling how swimming cells shape their dynamic environment, and vice-versa, and how collective effects emerge due to confinement and chemical fields, are interesting research directions.

**Advances in Science and Technology to Meet Challenges**

To meet the challenges outlined above, progress is required at various fronts: We first note that measuring three-dimensional swimming dynamics of agents in bulk still poses a major challenge, as it mostly relies on specialized equipment and is limited by low statistical accuracy. Data-analysis techniques, including tracking algorithms based on state-of-the-art machine-learning techniques and the recent framework of differential dynamic microscopy (DDM), may allow overcoming such limitations. To achieve a high statistical accuracy and enable such measurements in every laboratory having standard equipment, the use of DDM holds great potential. It relies, however, on both the processing of microscopy images and a fitting algorithm to appropriate models for swimming cells [10] and would so strongly benefit from open-source tools and new models. Additions to account for chemotactic behaviors will enable prediction of the chemotactic drift and relate it to molecular mechanisms of signalling pathways, a problem that is paramount in microbiology. Further, machine learning and related algorithms have the potential to improve current image processing techniques and deductions to study collective phenomena in more detail.

Along similar lines, models and simulations with varying degree of complexity and levels of coarse-graining have been established in recent years, which could be organized and made available to the community using open-source tools. While open-access simulation software, such as LAMMPS, is available to study primarily passive colloidal and polymer systems, extensions to active systems are sparse but could be of great value to the community. Moreover, combining tools of equilibrium polymer physics and hydrodynamics of filaments could be an excellent starting point for studying entangled active matter.

From an experimental perspective, microfluidic chips allow for the study of environmental complexities in a controlled manner. For example, establishing tools to maintain chemical gradients over long times, designing confined geometries that allow visualization of (three dimensional) transport inside them, and synthesizing complex fluids with well-characterized rheological properties are paramount for future progress. To identify relevant model systems, it will be important that physicists and engineers together with biologists, health experts, and material scientists systematically characterize microbial and other cellular habitats. This may reveal combinations of the aforementioned environmental complexities and will guide experiments and modelling. Manufacturing new synthetic (micro)swimmers and studying individual and collective responses may enable new opportunities in medicine, environmental sensing, and many more.

**Concluding Remarks**

The physics of motile agents in complex environments is a fascinating research area, with exemplars across various systems and scales. On the one hand the dynamics of motile agents are determined by both their biophysical (sensing) mechanisms and mechanical interactions with their surroundings, yet deciphering these contributions still remains a major challenge at both the single-particle and collective levels, calling for new theoretical modelling at different levels of coarse-graining. On the other hand, macroscopic phenomena, such as aggregation or large-scale flows, can be generated via interactions among individuals and their coupling to features of the external environment. An exciting opportunity in the future will be understanding the interdependence of fluid, elastic, and chemical fields. Continuing to grow synergies between life scientists, engineers, physicists, and applied mathematicians will allow us to develop new theoretical and experimental tools, including open-source software, to gain important insights.

**Acknowledgements**

H.A. Stone acknowledges the National Science Foundation for support from CBET-2127563 and the Princeton University Materials Research Science and Engineering Center, DMR-2011750.

## 2 -- Manipulation and assembly of passive objects by active particles

David Saintillan, University of California San Diego, USA

**Status**

Active particles and their suspensions, propelled by self-generated forces, have emerged as promising tools for various engineering applications requiring the complex assembly and actuation of passive objects. Their unique ability to convert ambient energy into either directed motion or microscopic stresses enables them to interact with their surroundings in remarkable ways, offering new paradigms for orchestrating the organization or manipulation of otherwise inert materials and structures at the microscale. Two main types of active systems have received attention and offer distinct types of capabilities: (i) dilute self-propelled autonomous microswimmers, either biological (swimming microorganisms) or synthetic (self-phoretic colloids), and (ii) active fluids composed of large collections of hydrodynamically interacting active agents, such as dense bacterial suspensions or mixtures of biological polymers and molecular motors. Each system presents its own set of promises and challenges.

Self-propelled active particles, including bacteria and self-phoretic colloids, perform correlated random walks in space, allowing them to transport payloads, push on surfaces, and generate coherent flows. Early instantiations of synthetic microswimmers were motivated by biomedical applications such as drug delivery: indeed, the ability of individual swimmers to transport various types of cargo has been demonstrated in dilute systems, and living microorganisms, in particular flagellated motile bacteria, are also capable of performing similar tasks. Active particles can also be harnessed to power the rotation of passive objects, especially those with chiral shapes such as 2D ratchets, which has led to the design of intricate micromechanical gears and motors powered by bacteria [1] (see Fig. 1). When interacting with multiple passive objects, e.g. a suspension of inert colloids, microswimmer collisions with the colloids can enhance their mixing and dispersion. They can also mediate long-ranged attractive forces between pairs of colloids [2], which are analogous to depletion interactions and have been used to promote colloidal assembly into microstructured gels whose morphologies are unlike those of passive gels [3] (see Fig. 2). Active colloids have been shown to have the potential to assemble into active architectures that can be externally programmed to perform a variety of motions [4]. They can also be used to drive shape reconfigurations of passive deformable objects.

Active extensile suspensions, above certain concentrations, display spontaneous flows that can also be utilized for actuating passive objects. In confined periodic channels, spontaneous unidirectional fluid motions can be generated and sustained, driven internally by the self-alignment of the active particles in the mean-field flow they generate [5]. Active suspensions can be used to power the translation or rotation of passive objects, e.g. the inner cylinder of a circular Couette device [6], or a suspended passive inclusion in the shape of a gear [7].

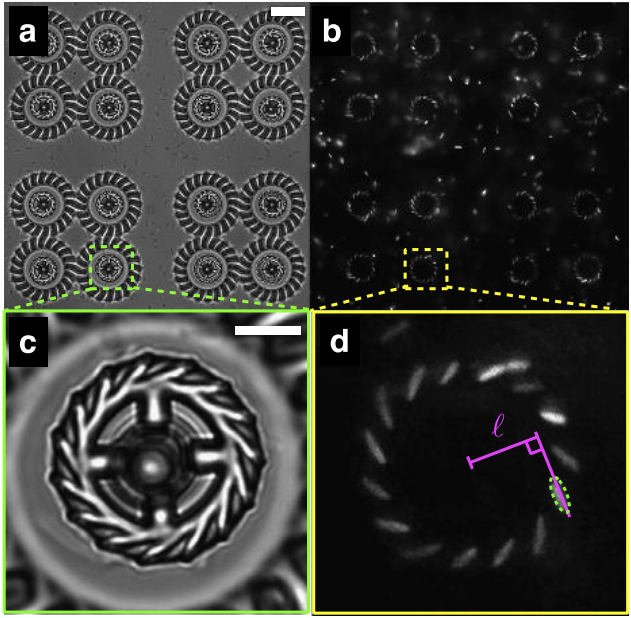

**Figure 1**. Light-controlled micromotors powered by bacteria: (a) array of 16 microfabricated rotors; (b) fluorescence image showing bacteria cell bodies, which tend to occupy microchambers distributed around the gears; (c,d) zoomed-in view of one of the rotors. Adapted from Ref. [1].

**Current and Future Challenges**

While the promise of active particles and their suspensions for designing novel methods for the assembly and manipulation of passive objects is clear, their full potential in applications has yet to be realized due to a variety of challenges, primarily associated with control. In applications involving targeted delivery of cargo by isolated microswimmers, control over the particle's trajectory is often desirable but can be tricky, due to thermal fluctuations among other factors. A wide range of strategies have been explored, including: external fields (magnetic, electric, light [1], acoustic), fluid flows, chemotaxis, or guiding by a passive liquid crystalline phase [8] or by rigid boundaries (flat, curved, or patterned). Designing methods for automated control of particle motion, e.g. based on feedback control algorithms, is an attractive avenue that has yet to be explored extensively in the case of self-propelled swimmers. Another limitation of current systems that has yet to be overcome is the fact that they are typically designed to perform one specific task. In some applications, one may wish to perform multiple distinct tasks in sequence (e.g. transport of building blocks, assembly, actuation) [4], which may require the coordinated action of several active entities with different properties, operating conditions, and control strategies. To date, the design of such multimodal platforms remains in its infancy and will require significant breakthroughs in both design and control. We also note that a vast majority of active matter systems relying on synthetic particles operate in two dimensions near flat substrates, and their deployment in 3D bulk systems presents significant hurdles.

The control of spontaneous flows arising in active suspensions is equally, if not more, challenging. Indeed, these flows are often the result of instabilities, display strong sensitivity on initial conditions and system perturbations, and can be unsteady and chaotic. Control strategies for active fluids are similar to those for dilute systems (e.g. external fields, flows, chemical gradients, geometry), but the precise effects of these controls on system dynamics are sometimes poorly understood or non-intuitive due to strong system nonlinearities.

An overarching challenge, both for systems involving isolated particles and for active fluids, is the incomplete understanding of the fundamental physics governing the system's dynamics. Much progress has been made over the last two decades on the theoretical and computational modeling of active matter systems, yet many models still rely on simplifying assumptions or idealized geometries that tend to limit their predicting capabilities.

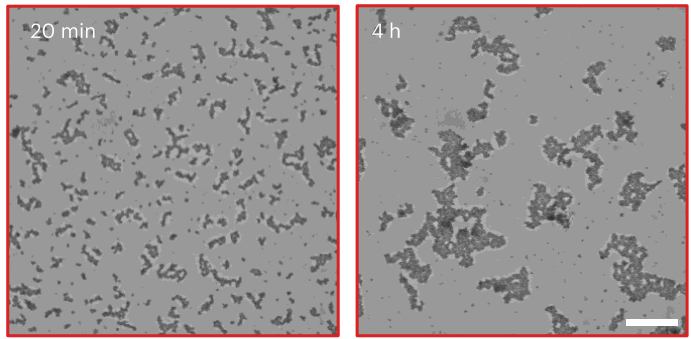

**Figure 2:** Aggregation dynamics in a monolayer of passive sticky colloids is accelerated by swimming *E. coli* bacteria and gives rise to aggregate morphologies that are unlike those observed in thermal systems. Reproduced from Ref. [3].

**Advances in Science and Technology to Meet Challenges**

Unleashing the full potential of active particles for manipulation and assembly will require advances in various areas. At the most basic level, progress in micro- and nanofabrication techniques should enable the design of novel active particles with specific properties and may open the door to new functionalities. Properties to be tuned or optimized include particle shape and buoyancy, Brownian diffusivity, swimming velocity, surface chemistry. In the case of biological swimmers, genetic engineering may prove a useful tool to tune particle properties such as flagella length and number or tumbling rate, and some achievements have already been made in this area in the case of bacteria. Particle design should also be guided by consideration of the ways particles may interact with their environment as well as with each other: exploiting complex physical effects (e.g. surface chemistry, fuel transport, interfacial and Marangoni stresses, interaction with light or thermal gradients, elasticity, interactions with boundaries etc.) has played a key role in many developments in the field and is likely to continue to offer novel ways of driving or controlling particle motions. Understanding and tuning interparticle interactions (steric, chemical, hydrodynamic) is also critical for applications involving the coordinated action or assembly of multiple active agents [4], and for the design of active suspensions displaying specific rheological behaviors and spontaneous flows [5].

As the complexity of experimental systems increases, the role of high-fidelity modeling will become ever more crucial. Accurate predictive multiphysics models (both analytical and computational) that are based on first principles and account for detailed experimental conditions would greatly benefit design strategies. Modern tools from optimal control theory as well as optimization algorithms will also be valuable. Finally, data-driven approaches including machine learning will undoubtedly play a sizeable role in future breakthroughs and can be applied in a variety of ways [9]: for the design of particles, experimental protocols, and control strategies, and more fundamentally for learning system dynamics and designing predictive models in situations where first-principles modeling is inadequate or incomplete [10].

**Concluding Remarks**

The ability to generate stresses and motion on the microscale is the hallmark of active matter systems, from self-propelled particles to active suspensions, and equips them with a unique ability to interact with, transport, and reshape passive objects and materials in their environment. This capability opens the door to a slew of novel methods for the manipulation and assembly of microscale objects, as has been demonstrated in wide variety of systems in recent years. The full potential of these methods for technological and biomedical applications remains to be achieved, and is likely to require concerted breakthroughs in microfabrication, soft matter physics, control theory and optimization, modeling and computation, as well as data science.

**Acknowledgements**

The author thanks Tanumoy Dhar, Jérémie Palacci and Daniel Grober for useful conversations, and acknowledges funding from National Science Foundation Grant No. CBET-1934199.

# 3 -- Perspectives on disordered active systems

Fernando Peruani, CY Cergy Paris University, France

**Status**

Most motile active matter studies -- including theoretical and experimental ones -- have focused on collections of actively moving particles through idealized, homogeneous environments. However, the great majority of active systems take place in nature, over a wide range of scales, in disordered, complex environments: active transport inside the cell, where the space is filled with organelles and vesicles, bacteria swimming in porous media such as in the soil or through complex environments such as the gastrointestinal tract, cancer cells penetrating the heterogeneous extracellular matrix, or herds migrating through forests and complex landscapes. We refer to this type of active systems, that are embedded in heterogenous, complex environments, as disordered active systems. For disordered active systems, external, time-dependent noise is not the only, and more importantly, not necessarily, the main source of fluctuations in the dynamics of the active particles. For instance, self-propelled particles moving on a substrate such as gliding bacteria or active colloidal rollers are strongly affected by substrate imperfections [1]. The motion of Eukaryotic cells is influenced by local roughness and heterogeneous adhesive properties of the substrate. The collective properties of Quincke rollers are dramatically affected by the presence of pinned obstacles [2]. In summary, in these examples, as in many other disordered active systems, quenched disorder, i.e. time-independent or frozen noise, is a key element that severely affects the large-scale statistical properties of these active systems. Among many other important effects, quenched disordered leads to jammed states [3, 4], spontaneous self-trapping of particles [5], and prevents, in two dimensions, the emergence of long-range order [6]. In short, the physics of disordered active systems is fundamentally different from the one of active systems in homogeneous media, and despite its paramount experimental relevance, remains largely unexplored. Moreover, understanding the physics of disordered active systems is essential to control and manipulate active matter via specifically designed prepatterned surfaces, obstacles arrays, or light fields. This is undoubtedly a promising area within active matter that, we anticipate, will quickly evolve in the following years. Investigating active systems in slowly evolving disordered fields, studying the impact of microswimmers motion on disordered environments and the feedback of the dynamics of the micro-swimmers on the environment, as well as exploring active glassy dynamics are all exciting research directions that emerge from the study of disordered active systems, which, we expect, will receive a great deal of attention in the coming years.

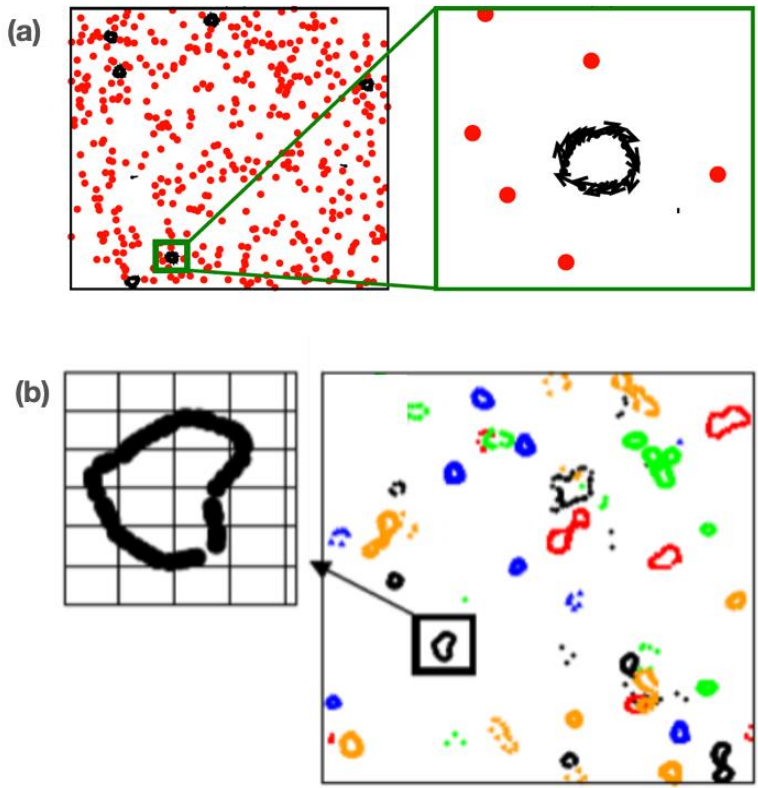

**Figure 1.** Simulations snapshots that display spontaneous trapping of active particles moving through (a) a random distribution of obstacles [from ref. [5]] and (b) a random stress field (b) [from ref. [6]].

**Current and Future Challenges**

Let us discuss the impact of quenched disorder – given priority to the pioneering works that appear on the subject -- first on scalar and then on vectorial active matter, i.e. SAM and VAM, respectively. In [3], it was found that Active Brownian Particles (ABPs) – i.e. active particles that do not experience torque when interacting among themselves or with obstacles – passing through a random distribution of pinned circular obstacles form clusters and become locally jammed. Specifically, ABPs nucleate around the pinned obstacles that thus act as a motility-induced phase separation (MIPS) seeds. If an external driving force is present, the generated flow exhibits a non-monotonic functional form with the active force, revealing that there exists an optimal activity level that maximizes the flow. Finally, in the limit of no dynamical noise (specifically, in the absence of angular fluctuations), the dynamics of the system freezes, and the system falls into an absorbing phase [4]. Except for ABPs, in general active particles experience torques when interacting with obstacles or among themselves (VAM). It was shown that torque interactions can generically lead to spontaneous particle trapping [5]. These traps, however, differ from the locally jammed configurations found in SAM: they correspond to closed orbits found by the active particles as they move through the random landscape of obstacles [5]; see Fig. 1(a). Importantly, it was found that due to these traps, active particles exhibit sub-diffusive behavior [5]. Another fundamental difference between active systems in homogeneous media and in the presence of quenched disorder, reported in [6], is that the self-organized polar order that spontaneously emerges in two-dimensions becomes quasi-long-ranged (QLRO) instead of long-ranged (LRO) as found in the foundational work of Vicsek et al. in 1995. Furthermore, it was also shown that that there exists an optimal (dynamic) noise that maximizes self-organized polar flows moving through disorder [6]. Moreover, it was found that the system becomes disordered at high as well as low (dynamical) noise values [6]. On the other hand, for active systems with metric-free interactions (i.e.

with Voronoi or Knn neighborhoods), the emergent order is long-ranged even in the presence of pinned obstacles. In this type of active systems, quenched disorder induces an effective density-order coupling that facilitates the emergence of traveling bands [8]. Counteractively, metric-free active matter in the presence of quenched disorder behaves as metric active matter in the absence of disorder [8].

We stress that a theoretical, analytical understanding of most of these results is missing (few exceptions are [5] and [7]). Moreover, disordered active systems remains poorly explored, numerically, analytically, and experimentally, and most findings are based on realization of quenched disorder based on random distributions of obstacles.

**Advances in Science and Technology to Meet Challenges**

It is possible to conceive more generic forms of quenched disorder than a random distribution of obstacles, some of which are amenable to analytical treatments and include generalizations that exhibit different symmetries; see Fig. 2. For ABPs, disorder can be expressed as a random potential field R(x) such that for the equation of motion for the i-th particle, in two-dimensions, takes the form:

$$\dot{x}_i = v_0 \hat{e}[\theta_i] - \mu \nabla_{x_i} U_{int} - \nabla_{x_i} R[x_i] + \hat{\eta}_i(t),$$

where $v_0$ is the active speed with $\hat{e}[\theta_i] = cos(\theta_i)\,\hat{x} + cos(\theta_i)\,\hat{y}$ the direction where it is applied, $U_{int}$ denotes an inter-particle interaction potential, $\mu$ the particle mobility, and $\hat{\eta}$ is a (dynamical) two-dimensional noise. The dynamics of $\theta_i$ for ABPs is either diffusive or given by Poissonian process as in run-and-tumble. For VAM, on the other hand, since particles do experience a torque, we must consider, in addition to the equation for $\dot{x}_i$, a non-trivial dynamic for $\theta_i$:

$$\dot{\theta}_i = -\mu_R\, \partial_{\theta_i} U_{int} - \partial_{\theta_i} R[x_i, \theta_i] + \xi_i(t),$$

where $\mu_R$ is the mobility associated to rotations and $\xi_i(t)$ is a (dynamical) noise. If $\partial_{\theta_i} R[x_i, \theta_i]$ is just a function of the position $x_i$, i.e. $\partial_{\theta_i} R[x_i, \theta_i] = f(x_i)$, then $f(x_i)$ represents a random torque field. Two other alternatives are to assume that $\partial_{\theta_i} R[x_i, \theta_i]$ defines either an aligning, random polar or random nematic, field, taking the form $\partial_{\theta_i} R[x_i, \theta_i] = f(x_i)\sin[\theta_i - \alpha(x_i)]$ (i.e. a random force field) and $\partial_{\theta_i} R[x_i, \theta_i] = f(x_i)\sin[2(\theta_i - \alpha(x_i))]$ (a random stress field), respectively. In the absence of dynamical noise, i.e. $\hat{\eta}_i(t) = 0$ and $\xi_i(t) = 0$ for all $t$, and additionally in the absence of interactions, the equations of motion corresponding to the random torque field exhibit a Hamiltonian structure, which prevents the existence of attractors [9]. In this scenario, the diffusion coefficient of particles can be estimated using the concept of shaking temperature [9]. On the other hand, for random force and random stress fields, or in the presence of particle interactions (for all random fields), the equations of motion display a dissipative structure. Note that a dissipative structure, implies that attractors are allowed. Thus, assuming the space is infinite, the existence of attractors is ensured and consequently the system dynamics falls asymptotically into a limit cycle; see Fig. 1(b). As results of this, spontaneous particle trapping occurs for all disordered active systems, except for the random torque field in the absence of interactions [9]. We note that these results were demonstrated in the limit of constant speed (see also [10]). And it remains to be proven whether, by adding dynamic noise, particle transport becomes subdiffusive. Extensions of this theoretical framework, including slowly varying random fields and/or coupling between particle motion and random fields, are possible and represent a promising research avenue.

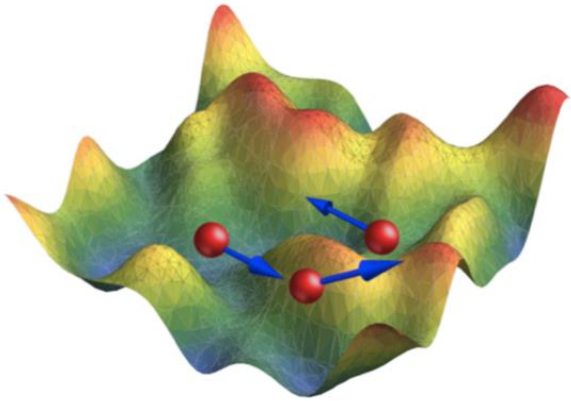

**Figure 2.** Illustration of active particles moving through a random field R. In Vectorial Active Matter, particles experience both, a random force and torque as they move through the random field R.

**Concluding Remarks**

Even though most natural active systems are undeniably disordered active systems, and despite recent findings that show unambiguously that the physics of disordered active systems is fundamentally different from the one of homogeneous active ones, these systems remain largely unexplored. Moreover, quenched disorder can be utilized to control active matter via the use of pre-patterned surfaces, arrays of obstacles, controllable complex liquid such as liquid crystals, or light fields, among many other examples. It is then essential to improve our understanding of the impact of quenched disorder on the dynamics of active systems beyond the results here discussed. Undoubtedly, the study of disordered active systems, given its theoretical and experimental relevance, will receive a great deal of attention.

**Acknowledgements**

F.P. acknowledges financial support from C. Y. Initiative of Excellence (grant "Investissements d'Avenir," Grant No. ANR-16-IDEX-0008); INEX 2021 Ambition Project CollInt; and Labex MME-DII, Projects No. 2021-258 and No. 2021-297.

# 4 – Active Cell-Mimicking Systems

Dmitry A. Fedosov, Thorsten Auth, and Gerhard Gompper,
Theoretical Physics of Living Matter, Institute for Advanced Simulation, Forschungszentrum Jülich, 52425 Jülich, Germany

**Status**

Biological cells can be considered as sophisticated soft micromachines with many internal mechanisms that provide mechanical stability and facilitate an active response to external cues and influences. The quest to understand these mechanisms has stimulated many attempts to construct artificial cell-mimicking systems – true to the statement of Richard Feynman "what I cannot create, I don't understand" in 1988. Furthermore, synthetic systems may one day provide new approaches to microrobotic systems and medical replacements.

First steps toward cell-mimicking systems have been made by incorporating internal force-generating components into giant lipid vesicles, and studying their out-of-equilibrium behavior. These components can be the same as in real cells -- such as microtubules, actin filaments, and motor proteins [1,2]. Alternatively, other components can be used, such as synthetic self-propelled particles (SPPs) [3,4], or biological microswimmers, such as bacteria [5,6]. We denote vesicles with enclosed active components as "active vesicles". Synthetic model systems have much less active components than biological cells, and thus their non-equilibrium behavior is easier to control and understand without a detailed knowledge of involved metabolic processes. Furthermore, a larger variety of materials can be used for their construction, extending their mechanical stability and functionality beyond existing biological systems. Examples include artificial membranes with favorable mechanical properties in comparison to biological membranes [7].

The generic model for active vesicles is a membrane bag with enclosed SPPs. Recent studies [3,4] have explored the behavior of such a system, where SPPs deform the surrounding vesicle membrane from inside. In the experiments, diffusiophoretic Janus colloids inside giant unilamellar vesicles (GUVs) are employed, and in the simulations, SPPs are modelled as active Brownian particles and the GUVs are represented by the dynamically-triangulated surface model. Figure 1 shows the simulated phase diagram of dynamic shapes of active vesicles as a function of the volume fraction $\phi$ of SPPs inside the vesicle and the Peclet number Pe = $v_p \sigma / D_t$, which characterizes the propulsion strength of the SPPs. Here, $v_p$ and $\sigma$ are the propulsion speed and diameter of the SPPs, and $D_t$ is their translational diffusion coefficient. The diagram shows a variety of highly dynamic, non-equilibrium vesicle shapes, including quasi-spherical vesicles with strongly enhanced "active" fluctuations, structures with multiple tether-like protrusions formed by individual SPPs or small clusters, and prolate and bola-like configurations. For the understanding of this behavior, the competition of membrane-bending forces, active pushing forces from SPPs and SPP-clusters, membrane-mediated interactions between SPPs, and the passive and active membrane tension must be considered [3,4].

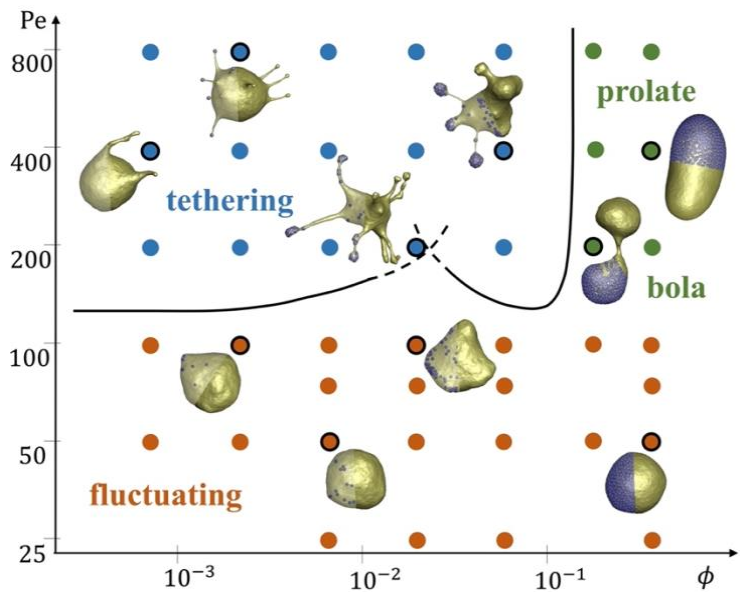

**Figure 1.** State diagram of active vesicles as a function of Peclet number Pe and the volume fraction $\phi$ of enclosed SPPs. Three regimes are indicated: tethering (blue), fluctuating (red), and bola/prolate (green) vesicle shapes. Snapshots are displayed for the points marked by open black circles. Theoretical estimates of the critical $Pe_{teth}$ for tether formation are also shown (black lines). Adapted with permission from Ref. [3].

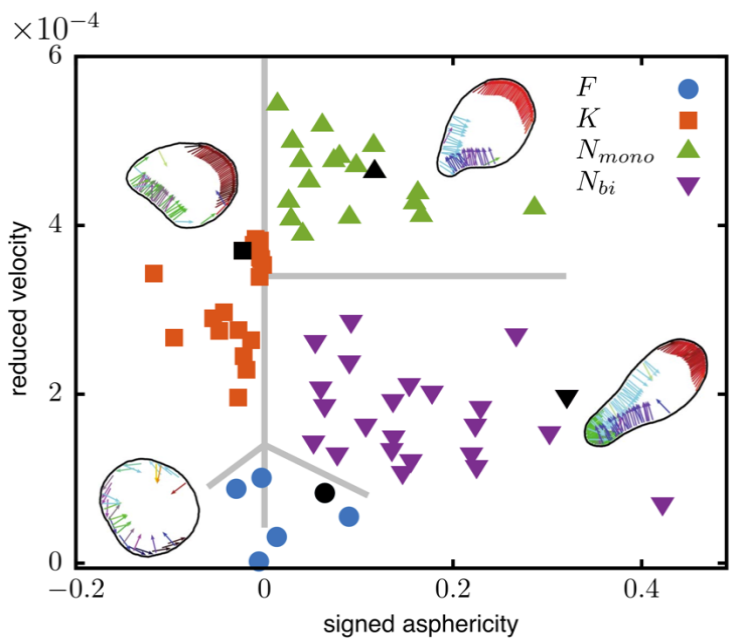

**Figure 2.** Asphericity and reduced velocities of 2D active vesicles with enclosed self-propelled filaments. The filaments are attached to vesicle membrane and can pull or push on it. Fluctuating (F), keratocyte-like (K), and neutrophil-like (N) vesicles are shown for various membrane properties, filament self-propulsion forces, and substrate frictions on filaments. The simulation snapshots show the vesicle shapes that correspond to the black symbols. Adapted with permission from Ref. [8].

When the vesicle adheres to a substrate, and the active particles are not spherical but elongated, directed motility of the active vesicle can be observed, because momentum can be transferred to the substrate. With the ability of spontaneous symmetry breaking, elongated self-propelled particles can attain polar alignment, and induce a net propulsion force, which leads to persistent directed motion of the vesicle, see Fig. 2. Two-dimensional model systems show complex motion on a patterned substrate with varying friction forces or passivated areas [8,9]. The observed response of the active vesicles to the external stimuli hints to a complex mechanical information processing by internal self-organization. For example, the internal rearrangement of self-propelled filaments in active vesicles can lead to either reflection at or sticking to boundaries. The reflection angle at planar boundaries is predicted to be independent of the incident angle and appears to be defined by the internal processes only, such as the dominance of pushing or pulling forces exerted on the membrane, in qualitative agreement with experimental observations for motile biological cells.

**Current and Future Challenges**
A closer look at the tether formation for a system shown in Fig. 1 reveals that this process is indeed quite difficult to realize with hydrodynamic self-propulsion forces. The extension of a tether would require both membrane and fluid flow into the tether, but a microswimmer at the front end is actually pushing fluid to the back [10]. Furthermore, simple force balance at the front end indicates that no pulling forces can be generated. Then, how do Janus particles pull tethers? A possible explanation could be related to osmotic-pressure differences and fluid permeability through the membrane. Furthermore, vesicles with internal force generation cannot display persistent translational motion without a substrate. Thus, forces have to be generated on the outside, for example by attaching active particles to the outside of the vesicle membrane. How can such a machine be realised experimentally?

Since recent developments in microscale and nanoscale technologies allow a well-controlled construction of cell-like compartments with cell-based materials, artificial bio-mimicking systems should in principle allow the step-by-step elucidation of the physical principles that govern cellular multiscale self-organization. One of the major challenges is to connect macroscopic behavior of such systems to mesoscopic self-organization and properties of internal components. *In-silico* models mimicking experimental systems can potentially bridge this gap, by providing answers to which mechanisms and parameters of the system are key to its macroscopic behavior.

An important and probably far-reaching goal is to equip bottom-up synthetic systems and in-silico models with the ability to mechanically and/or chemically sense their environment, so that they can react and adopt to changing geometries or existing chemical gradients. These sensing mechanisms have to be incorporated into internal structures of the membrane, for example, by the lipid composition or concentration of membrane proteins, which then transfers the information to the force-generating internal components. How can such active systems be designed to mimic various aspects of information processing by biological cells?

With the development of advanced active systems, it is plausible to expect their usage in various applications, such as microrobots in living organisms. Here, an interesting question is whether such synthetic systems can be made to be biocompatible and can in some respects surpass the properties of biological cells established by evolution.

**Advances in Science and Technology to Meet Challenges**
In addition to experimental realizations of cell-mimicking systems and active vesicles, theoretical and simulation approaches provide an important contribution for their understanding and design. *In-silico* approaches and multi-scale modeling permit detailed dynamic measurements of local chemical and mechanical conditions which can be associated with the global behavior of synthetic systems. Furthermore, computer simulations permit a direct and systematic exploration of the importance of various mesoscopic characteristics (e.g. mechanical characteristics of active components and membrane, and their interactions) for the self-organization of functional active systems. Here, a close collaboration between experiments and modeling is necessary.

Hydrodynamic flow is an intriguing mechanism in biological cells that may contribute to several active processes, such as the cytoplasmic motion of the cytoskeleton. To date, models that contain active mechanical components often disregard hydrodynamics. Using exascale computing, systematic studies of the behavior of active cell-mimicking systems that contain hydrodynamic interactions between various internal components appear feasible in the nearer future.

**Concluding Remarks**

Artificial active-vesicle systems can aid in the elucidation of physical mechanisms that govern cellular behavior. Furthermore, they can be considered as soft micro-robots, which can autonomously preform various functions. The ability to design and construct soft micro-robots with combined activity and adaptability has the potential to revolutionize the fields of soft active materials and nanomedicine (e.g., targeted drug-delivery systems).

**Acknowledgements**

Many stimulating discussions and collaborations with Clara Abaurrea-Velasco, Masoud Hoore, Priyanka Iyer, Florian Overberg, Cesar Rodriguez-Emmenegger, Jan Vermant, and Hanumantha Rao Vutukuri, as well as support by the EU ETN "Physics of Microbial Motility" (PHYMOT), are gratefully acknowledged.

# 5 -- Chemically propelled particles from complex interactions to complex dynamics

Cécile Cottin-Bizonne and Christophe Ybert
Université de Lyon, Université Claude Bernard Lyon 1, CNRS, Institut Lumière Matière, Villeurbanne, France

**Status**
Externally-powered active matter using electric, magnetic or optical fields, has developed as a powerful and versatile model experimental platform for exploring the physics of active matter [1]. Yet the richest examples of active matter dynamics and organization are provided by biological systems whose power originates from harvesting chemical energy from their environment. There, chemical fields not only provide the energy source, but also act as a ubiquitous signaling pathway crucial for various biological functions. They can trigger biased self-propelled dynamics, such as chemotaxis or aerotaxis, guiding organisms towards favorable ecological niches. They can also serve as an interaction cue between individuals in quorum sensing phenomena or induce transition in the organisms' state and motile lifestyle, such as the formation of biofilms. In that respect, model chemically-propelled systems constitute a masterpiece of active matter investigations to decipher the intricate interplay between activity, motion, and chemical interactions. They can be viewed as a physicist tool to rationalize biological phenomena, retaining key features without the complexity of living systems. They also constitute a bioinspired route towards synthetic systems with augmented properties. Various experimental realizations, including catalytic particles, self-phoretic particles, or self-propelled Marangoni droplets, have been explored, allowing for the emission and reaction to chemical fields. These particles' individual motion and their behavior in semi-dilute regimes, where interactions often promote collective dynamics and cluster formation have been extensively investigated. Yet except maybe for the swimming droplets [2], [3], the physics of chemical swimmers has not always been elucidated beyond a coarse level while the details of even semi-dilute properties appear sensitive to the fine characteristics of the systems. Indeed, experimentally, the underlying chemical field has only recently been characterized for droplets and otherwise remains as a blind spot of many systems. This becomes even more critical when considering denser crowded systems, obstacles or confining boundaries, which all impact the characteristics of the chemical environment.

To further advance on these topics, it is now crucial to develop an expanded description for chemically powered systems. This description should not only capture the complexity of chemical interactions but also account for a wider range of effects such as confinement that may be compliant, inducing novel feedback loops. This step is essential for a more accurate description of the behavior of biological micro-swimmers and to unveil novel physical phenomena. Additionally, this also involves extending the features of synthetic systems to incorporate more of the characteristics observed in biosystems, thus enhancing their level of interactions.

**Current and Future Challenges**
Investigations on chemically-powered active particles will now expand toward increasing complexity, beyond the previous dilute systems or the study of spontaneous collective dynamics at intermediate densities without constraints. One direction corresponds to active materials, be it dense active systems or active particles embedded in a supporting matrix, and to their dynamical and mechanical responses. Experimental focus has been so far on materials composed of externally powered active spinners. However, exploring the mechanical properties of active solids constructed from chemically propelled particles emitting individual chemical fields and interacting with one another remains intriguing. So far, active glasses have been explored from an aging perspective [4], but much remains to be done regarding their mechanical properties. Similarly, how this chemical coupling competes with both hydrodynamics and elastic properties of the environment is a largely open question of fundamental interest and direct relevance to many biological environments.

Pushing forward this line of interaction with complex environments, a second future challenge arises from confinement's influence on active systems. While hard confinement has been explored, especially through hydrodynamic interactions (single wall, pillars, porous media), soft and compliant confinement remains relatively unexplored but promises novel insights as shown in simple cases [5]. Moreover, if we think about biological systems, the environment itself hardly behaves as a frozen field and often incorporates its own amount of activity. Interactions between two active systems can occur due to steric confinement or chemical signatures of their activities. In this respect, fluid-fluid interfaces have already demonstrated interesting specific effects associated with their ability to respond to the chemical heterogeneities produced by active particles or surfers [6].

Finally, a third challenging direction for the future may come from the characteristics of the chemically-propelled particles themselves. So far they have been essentially restricted to monodisperse swimming objects interacting via their common fuel and waste chemical fields. The heterogeneity and/or polydispersity in particle characteristics have little been explored. This would further increase asymmetry of interactions, in addition with the interest of non-reciprocality which has recently emerged [7]. In this respect, one interesting route would be to increase the level of chemical interaction beyond the present single fuel/waste species paradigm. Inspired by molecular active systems [8], an appealing approach could take the form of mixtures of particles coupled through catalytic cycles -where one particle type produces waste compounds acting as fuel for another particle type-.

**Advances in Science and Technology to Meet Challenges**
To address previous challenges, various bottlenecks must be tackled: either related to experimental tools and characterization techniques, to methods and theoretical approaches or to systems.

As surprising as it may, experimental in-depth characterization of chemically-propelled particles are scarce. Their phoretic response to fuel and waste chemicals often lacks independent and direct measurements that would help constrain the different theoretical regimes. This is even more true for interactions where a strong need exists for direct investigations. Even pair interactions are actually challenging, as the phase space contains distance, relative orientations and may incorporate history effects. In practice, this requires achieving some control on the initial release of particles to build statistics, though development of inference methods could offer an alternative approach. More generally, direct information on chemical gradients around self-propelled particles, potentially using fluorescent probes targeting popular fuel/waste compounds (hydrogen peroxide or oxygen) or pH heterogeneity are needed to deepen our understanding of these systems. Simultaneously, characterizing macroscopic properties, like rheological behavior [9], in dense assemblies of active units or gels embedded with active particles, is a crucial missing step for developing active solids with novel properties.

To date, a large amount of the statistical physics approaches, including advances towards mechanistically important aspects such as active pressure, eventually falls back to dry active matter, while hydrodynamic interactions were accounted for in extensions from rheological approaches of colloidal suspension. In complex situations, accurate descriptions of chemically-propelled particles should inherently consider both hydrodynamic and chemical couplings. A minimal reference model is still to be developed, such as Active Brownian or run-and-tumble particles have been so far. Note that one additional complexity may come from memory effects from trails left by active particles [2]. Describing active particles at a fluid interface is even more complex due to stronger couplings.

Finally, among the mentioned challenges, some require expanding the capabilities of chemically-propelled particles. In this respect, active droplets show promise for further development, with achievements like the ability to change the swimming regime in response to a chemical cue- or asymmetric interaction involving multiple chemicals [10]. Developing active solid particles, possibly via 3D printing, with varied shapes and multiple reactive coatings also holds potential. Moreover, a growing interest develops for liquid-liquid phase separation leading to membraneless intra-cell compartments which concentrate enzymes and exhibit strong reactivity. Experimental development in this area could provide remarkable tools for self-propelled active matter.

**Concluding Remarks**

Chemically-propelled colloids have been seminal as a model active system to study dilute or semi-dilute regimes, although they have since been largely complemented for instance by systems with external power sources. Yet, they intrinsically bear unique interaction characteristics underlying the remarkable dynamical and self-organizing properties of biological systems, and thus should constitute a cornerstone for future investigations of their underlying physical mechanisms. To do so, it will be necessary to consider the complexity of chemical interactions and potential feedback effects. Promising avenues for future studies encompass dense assemblies of active colloids forming novel active materials with unique properties, systems interacting with soft and compliant confinement environments, or examining the effects of particle polydispersity. Overcoming associated challenges, such as elucidating chemical gradients, characterizing the mechanical properties of active unit assemblies, and developing comprehensive models incorporating chemical and hydrodynamic couplings, will be crucial. Ultimately, these developments will also build on the elaboration of novel systems of increased capability as compared to the present prototypal system of monodisperse particles with identical fixed activity. We anticipate that this may benefit from -and possibly contribute- to the development of protocells which provide minimal artificial systems with an increasing number of biomimetic capabilities.

**Acknowledgements**
We acknowledge financial support from ANR-20-CE30-0034 BACMAG and ETN-PHYMOT within the European Union's Horizon 2020 research and innovation programme under the Marie Sklodowska-Curie grant agreement No 955910. We thank Nicolas Bain, Hélène Delanoë-Ayari, François Detcheverry and Mathieu Leocmach, for fruitful discussions and collaborations on the topics discussed here.

# 6 -- Lagrangian tracking: unveiling the behavioral variability of micro-organisms in changing environments

Eric Clément[1,2], Thierry Darnige[1], and Anke Lindner[1],
[1] Laboratoire PMMH-ESPCI, UMR 7636 CNRS-PSL-Research University, Sorbonne Université, Université Paris Cité, 75005 Paris, France.
[2] Institut Universitaire de France, Paris, France

**Status**
Microorganisms are ubiquitous on earth. They develop multiple forms of collective and symbiotic interactions, crucial to maintaining life on our planet and ensuring a sustainable environment for human species. Many of those microorganisms possess the ability to move and orient their motion in response to environmental information, allowing them to find their ecological niche or to escape from danger. Clarifying the link between motility and environment is crucial to capture large-scale transport or contamination processes but also plays a central role in the emergence of collective organization phenomena. The inherent complexity of the problem stems from the great diversity of confining structures explored by microorganisms, leading to a large variety of chemical, photonic, or mechanical signals perceived. In return, these signals lead to variations and adaptation of internal molecular variables influencing micro-organism mobility, in the presence or absence of driving flows. Gaining fundamental understanding of the role of self-propulsion and adapted motility for micro-organisms navigating through heterogeneous environments and ever-changing situations, can be gained on the basis of quantitative experimental results. Such an endeavour requires not only well controlled experimental model environments, but also sophisticated observation techniques suited to study self-propelled micro-organisms in complex and often dense surroundings.

Recent advances in microfluidic techniques have propelled the development of model systems with increasingly well-controlled environmental conditions, featuring tailored geometries and precise control over chemical, lighting, and flow conditions. These setups not only attempt to replicate salient features found in nature, but also serve as innovative experimental platforms for exploring emerging concepts in active matter. Concurrently, significant efforts have been made to directly visualize, under the microscope, the individual and collective dynamics of microorganisms within such environments. Two complementary visions are known from hydrodynamics. In an Eulerian approach, fields, as density, velocity or concentration, are observed in the laboratory frame in fixed volumes. In a Lagrangian approach one follows a fluid element or in the case of a motile micro-organism the trajectory of the latter. Quantities such as concentrations or velocity gradients are evaluated locally, in the reference frame of the micro-organism. For practical reasons many measurements are performed and expressed as Eulerian fields, however transport properties or dispersion processes are best viewed in a Lagrangian framework. Both approaches face specific challenges due to the optical complexity of the heterogeneous environments and the inherent mobility of the micro-organism and will be discussed in the next sections.

**Current and Future Challenges**
We will mostly focus on techniques developed to visualize motile bacteria such as E-coli, but the described techniques are not restricted to the latter. Assessing large numbers of bacterial trajectories in an Eulerian framework is relatively easy nowadays, and relies either on fluorescent staining or phase contrast methods readily available on modern microscopes. Such observations represent, however, projections of short tracks in the 2D observation plane. The surrounding hydrodynamic flow fields can be obtained precisely using micro-PIV techniques. Recent advancements in sophisticated holographic techniques have extended the boundaries of Eulerian observations, enabling the simultaneous capture of hundreds of bacterial tracks in three dimensions [2]. Such observations lead to the discovery of bacterial rheotaxis, motility strategies [2] or wall entrapment [1,3]. These Eulerian techniques find their limits in the presence of a fixed frame of observation, strongly limiting the observation times. Individual microorganisms when swimming or transported by a flow can readily leave or enter the field of observation. To obtain a clear vision on individual time-changing properties manifested in the evolution of swimming velocities, tumbling rates, adaptation to chemical or lighting history, a Lagrangian approach is more appropriate.

One of the pioneers to recognize the necessity of a Lagrangian perspective was Howard Berg in the 1970s, marking a significant leap in comprehending the individual behavior of bacteria. In a remarkable achievement for his time, Berg constructed a 3D Lagrangian tracking device capable of monitoring motile E. coli for tenths of seconds [4].

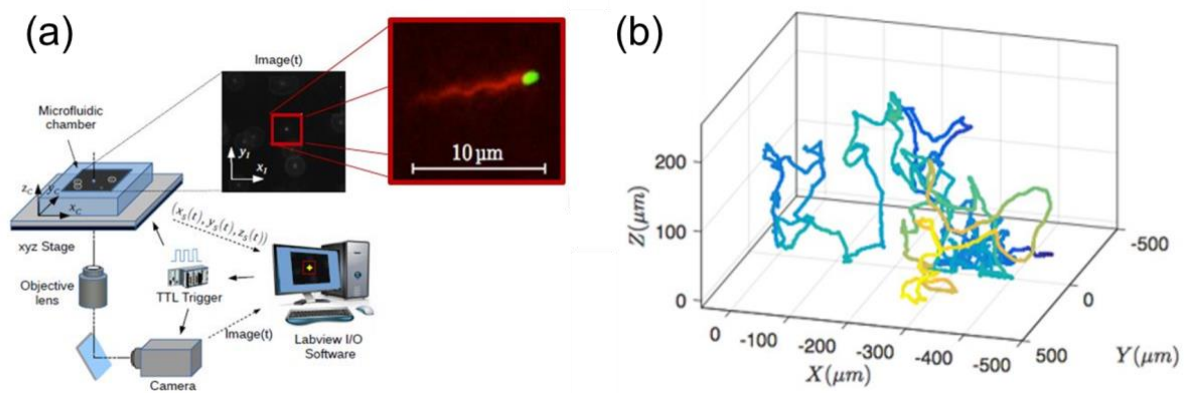

Figure 1. Lagrangian 3D tracking methods. (a) Sketch of the tracking apparatus [1] under the microscope showing a 3D (XYZ) stage driven by a computerized algorithm taking an image (80 FPS) around a tagged object within a virtual trapping area (inset). The program analyses the image and provides information to the stage such as to keep the object in the central part of the trapping area and in the focal plane. Large inset: magnification of a tracked object (a motile E.coli) when the system is working in the "two-colors mode" [1]. Tracking is performed on the green fluorescent body, but in parallel the flagella bundle is visualized in red fluorescent emission. (b) Example of a 500 seconds track of an E.coli exploring the space between two glass slides. Colour code corresponds to Y-direction depth (Courtesy of Renaud Baillou, PMMH-ESPCI, PhD Thesis, 2023).

This groundbreaking apparatus revealed the kinematics of multi-flagellated microorganisms as a combination of runs and tumbles. Later the importance of the flagella bundling time was identified using a similar method [5]. The influence of internal protein-based behavioral processes, changing slowly in time, was subsequently identified using a refined set-up with the ability to track individual bacteria for several hundreds of seconds [6]. These tracking methods rely on a moving microscope stage, keeping the bacteria in the image and in focus (Fig. 1). The 3D movement is obtained through a feedback loop and image analysis. Tracking individual bacteria for extended periods enabled the differentiation between phenotypical variability within the population and behavioral variability exhibited by individual organisms [6]. Lagrangian methods were also used with success to address questions on bacterial dispersion in the presence of a transporting flow or persistent motion driven by chemo- or phototaxis. A fundamental constraint of Lagrangian methods is the practical limitation on the number of tracks that can be achieved, rendering it complementary to Eulerian techniques.

**Advances in Science and Technology to Meet Challenges**
Despite significant advances made in micro-organism tracking in recent years, establishing a direct link between motility and environment remains challenging. This challenge arises, in part, from optically complex environments that can be crowded or feature multiple, sometimes curved, interfaces between different materials, rendering precise 3D tracking impossible. However, such precision in tracking is essential for obtaining accurate information about the properties of bacterial surroundings or for detecting internal signalling processes.

The following future developments will help improve tracking performance. A new category of algorithms using artificial intelligence (AI) has demonstrated effectiveness in accurately identifying the positions of microorganisms, especially in relation to the focal plane [7]. This ensures precise maintenance of microorganisms within the focal plane, and the improved focusing quality introduces new possibilities, such as the capability to capture reporter fluorescent signals originating from internal molecular processes [8]. Such analysis could provide a list of relevant parameters defining the time resolved status of the microorganism. Fluorescent signals can also be leveraged to identify local concentrations of specific chemicals, for instance, of oxygen [9]. AI techniques can be readily adapted to various types of lighting, including fluorescence, phase-contrast, bright or dark field, and thus applied across diverse microorganisms. Moreover, preliminary results indicate that algorithms, which include a training phase, remain effective in optically complex biological or medically relevant media, such as biogels or mucus extracts. The integration of Lagrangian tracking and Digital Holography Microscopy (DHM) [10] holds the potential for time-resolved 3D visualization of a tracked object and its immediate surroundings. Presently, DHM computing algorithms, capable of analyzing intensity and phase signals, facilitate rapid 3D reconstruction of living biological cells at a given instant. Conducting DHM within a confined moving volume, tracking a specific cell, introduces novel analytical opportunities, including the determination of the cell's 3D orientation in relation to its immediate environment, encompassing other microorganisms, prey, and the geometry of confining media. To enhance efficiency, trained AI algorithms could be employed for automated analysis, hence providing precise and time-resolved information on bacterial motility with respect to their surroundings.

Harnessing the advancements in Lagrangian techniques for assessing both internal and external parameters will shed light on the learning capabilities of microorganisms in the face of complex and dynamically changing environments. Furthermore, this approach holds future potential for delivering direct feedback on the motility of microorganisms, influencing their exploration processes, by using for example light signals.

**Concluding Remarks**
The future evolution of Lagrangian methods will provide time-resolved insights into both the immediate surroundings of a microorganism in motion and its internal molecular state while navigating environments of varying complexity. The prospect involves the development of an innovative Lagrangian toolbox, stemming from either technical advancements or conceptual breakthroughs inspired by existing techniques. The goal is not solely to enhance 3D tracking performance and versatility, applicable across various microorganisms, but also to pave the way to new inquiries. This includes addressing questions related to optimal exploration paths, learning capabilities, and ultimately, the potential to provide direct feedback to the tracked microorganisms, thereby influencing and controlling their exploration. This opens up a new avenue for investigating various taxis (phototaxis, aerotaxis, chemotaxis, magnetotaxis, and rheotaxis) and gaining direct insights into microorganism behavior and adaptation that were not previously attainable. A central question involves quantifying the impact of confinement, geometrical disorder, complex medium rheology, and flow. This approach promises new insights into the connections between internal sensory circuitry and motility, providing a fresh perspective on adaptation processes.

**Acknowledgements**
We acknowledge the financial support of the ANR-22-CE30 grant "Push-pull", the Innovative Training Network H2020-MSCA-ITN-2020 - PHYMOT "Physics of Microbial Motility" and the ERC Consolidator Grant PaDyFlow (Agreement 682367). This work has received the support of Institut Pierre-Gilles de Gennes (Équipement d'Excellence, "Investissements d'avenir", program ANR-10- EQPX-34). and E.C.

is supported by the Institut Universitaire de France. We thank Profs. Giovanni Volpe and Daniel Midtved for enlightening discussions and collaboration on AI methodology developments.

# 7 -- Chiral Microswimmers: From Hydrodynamic Bound States to Phototaxis

Raymond E Goldstein, Department of Applied Mathematics and Theoretical Physics, University of Cambridge, Cambridge, UK

**Status**

One of the most fundamental questions in biology concerns the driving forces that led to evolutionary transitions from unicellular to multicellular life. This transformation is not simply a matter of size, but also of biological "complexity", typically viewed in terms of the division of labour within an organism into specialized cell types. While this area has been of interest to biologists for well over a century, over the past two decades there has been growing interest among physical scientists in this question, as life at the microscale is dominated by effects such as diffusion, buoyancy, mixing and sensing.

Among the most appropriate model organisms to study transitions to multicellularity are green algae [1], eukaryotes that span from unicellular species ten microns in size to multicellular ones several millimetres across with up to 50,000 cells. The multicellular forms are planar or spheroidal, with cells embedded in an extracellular matrix (ECM) they export. Their motility is driven by the undulations of flagella whose microstructure is among the most highly conserved features in biology. As these organisms are photosynthetic, they have evolved a form of locomotion that is tightly linked to a sensory pathway for light based on a directional photosensor in each cell; each species spins about a body-fixed axis as its flagella beat, such that when the body axis is misaligned with a light source the photoreceptors experience a periodic signal.

Among the many discoveries in biological fluid dynamics that have arisen from studies of green algae, one of the most striking is the formation of "hydrodynamic bound states" when pairs of *Volvox* swim near a solid surface, are attracted together and orbit each other (Fig. 1). Consistent with an analysis by Squires [3], the attraction arises from the downward Stokeslet flows arising from the negative buoyancy of the colonies. At the same time, hydrodynamic interactions between spinning colonies lead to changes in the spinning speed as a function of distance – a kind of "spin-orbit" coupling. This was perhaps the first quantified example of "non-reciprocal interactions" between chiral microswimmers. Rediscovered multiple times since using bacteria [4], starfish embryos [5] and optically driven colloids [6], these interactions are now understood to lead to unusual elastic properties of very large arrays of chiral microswimmers [5] under the heading of "odd elasticity". The ubiquity and linkage of chiral swimming and taxis in the microscopic world highlights the intrinsic interest in these phenomena.

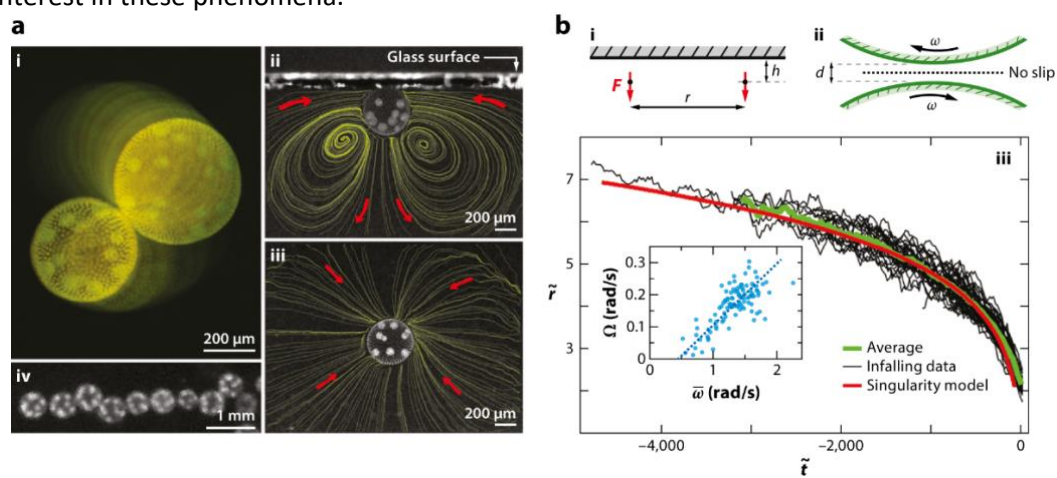

*Figure 3. Hydrodynamic bound states of the green algae Volvox carteri [2]. (a) Views from above and the side of colonies, showing stokeslet-driven flows and formation of a linear array. (b) Geometry of downward Stokeslets, interactions of spinning colonies, infalling trajectories – scaled colony separation versus scaled time, and orbital frequency versus spinning frequency (inset). From [1] with permission.*

**Current and Future Challenges**

The architecture and locomotion of chiral microswimmers provide several research challenges that are the subject of current work. The first involves a general question: How do cells build structures external to themselves in a robust and accurate manner? Here the specific question concerns the ECM of these algae, which is exported in such a manner that the growing colony remains spherical. Recent work has shown by Voronoi tessellations of the surface of *Volvox* that the neighbourhood areas around each somatic cell are very broadly distributed, having a functional form consistent with a gamma distribution, and that same form is found for the Voronoi volume distribution around cells in the lab-evolved multicellular organism ``snowflake yeast'' [7]. For *Volvox*, this finding reveals a hitherto unknown type of biological noise associated with ECM production and raises the question of whether these fluctuations away from uniformity are a consequence of noisy ECM production at the level of individual somatic cells, or a consequence of feedback processes at work. Moreover, the chiral swimming that lies at the heart of non-reciprocal interactions and phototaxis raises profound questions in development. What features of cell division, embryonic inversion [8] and ECM production introduce and preserve the chiral arrangement of flagellar beat planes during development? What is the frequency and spatial distribution of the topological "defects" that are found in Voronoi tessellations of the ECM (Fig. 2)? Moving beyond the simplest of multicellular organisms, I suggest also examining sponges, the basalmost animals, whose complex microfluidic architecture of flagellated filter-feeding cells in spherical chambers embedded in an ECM is a paradigm for the generation complex network architecture external to cells.

A second issue concerns phototaxis; work on several species of green algae [9] and larvae of a marine invertebrate has shown that there is a shared "adaptive" response of the flagellar apparatus to the periodical light signals received by the photosensors on phototactic organisms. Over a wide range of species this response is tuned to the spinning frequency of the organism itself in such a manner as to optimise the turning rate toward the light. At present it is unknown what types of genetic networks underlie adaptive phototaxis and what might have been the evolutionary origins of those structures, which basically act as a bandpass filter. The phototactic system may be an ideal one to understand a sophisticated example of cellular decision-making in the presence of noisy signals.

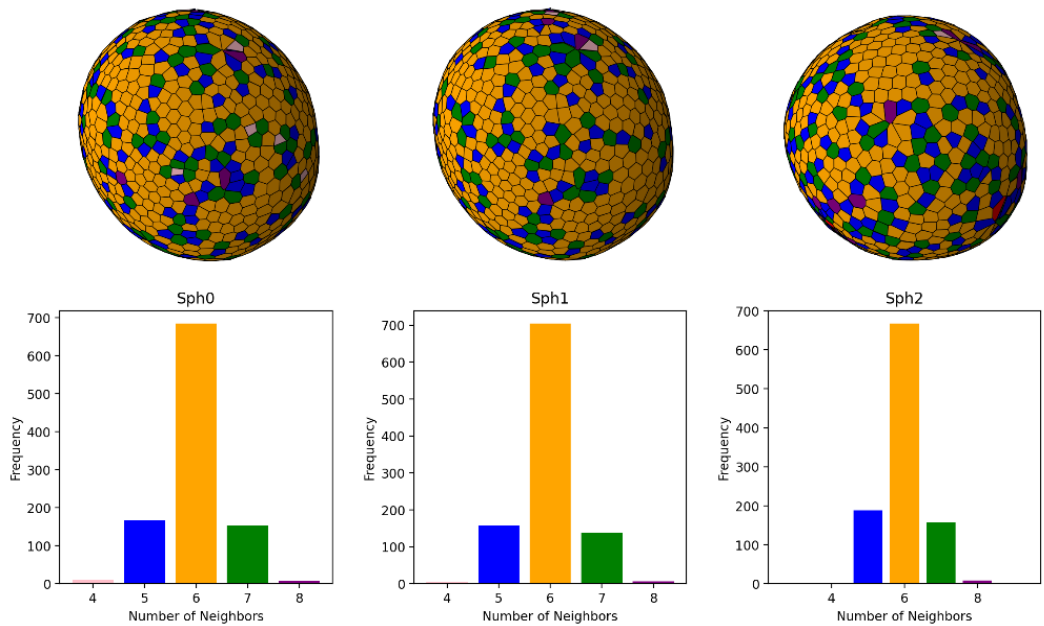

*Figure 4. Distribution of topological defects in the cellular neighbourhoods of three colonies of the multicellular alga Volvox carteri, obtained by light-sheet microscopy [7]. Figure courtesy of A. Srinivasan.*

**Advances in Science and Technology to Meet Challenges**

The challenges outlined above can best be met by a holistic approach that involves methods from developmental biology, fluid dynamics, statistical physics, and nonlinear dynamics.  For the general problem of understanding ECM generation, with implications for cellular positioning and orientation, it is necessary to measure the growth of organisms over days, with submicron precision in cellular tracking, all with the diurnal temperature and light conditions associated with normal physiology.  We can anticipate that this will be achieved by suitable developments in light-sheet imaging, advanced image processing methods for accurate image segmentation, and advances in fluorescent labelling of ECM components.   The theoretical issue at hand is best described as understanding the geometric features of "stochastic Voronoi tessellations" and how they relate to underlying processes such as export of ECM components.

The linkage between tuned adaptive phototactic flagellar responses and organism spinning will likely benefit from careful studies using directed evolution, where phenotypic changes arising from selection for aspects of tactic behaviour may give clues to the processes at work.  At the single-cell level we anticipate that various ohmic technologies will be important in understanding at a systems biology level the origins and structure of networks that allow for adaptive phototaxis and its connections to adaptation in other contexts, such as bacterial chemotaxis.

Finally, if we are to understanding the development and physiology of topologically complex microfluidic networks in basal animals such as sponges it is likely that new growth and imaging methods will need to be developed due to the opacity of these organisms.  For example, is it possible to coerce sponges to grow quasi-two-dimensionally?  Can methods such as optical coherence tomography be used to quantify the growth processes?  The most important theoretical issue is the development of a quantitative understanding of how an object that begins life in the embryonic stage with genus zero can end up with a structured network having a genus measured in the hundreds or thousands, with precise sorting of cells and ECM as observed in nature.  Beyond this are issues of emergent tissue-level behaviour such as global contractions [10], which have only begun to be investigated.

**Concluding Remarks**

The brief overview above highlights the important role that chiral microswimmers have played in various areas in biological physics and the kinds of profound questions that their biology poses for the physical sciences in general.  I expect that the most interesting developments will come from unexpected connections between disparate areas in science, whether from problems in pure mathematics associated with stochastic partitioning of space, or the nonlinear dynamic of decision-making.  To be successful, the experimental mindset must also move beyond the constraints of any one viewpoint such as fluid dynamics or genetics and embrace a more holistic approach to deciphering the workings of life.

**Acknowledgements**

*I thank Anand Srinivasan for the data in Fig. 2, Kyriacos Leptos for numerous discussions and am grateful for support from Wellcome Trust Investigator Grant 207510/Z/17/Z, The Marine Microbiology Initiative of the Gordon and Betty Moore Foundation (Grant 7523), and the John Templeton Foundation.*

## 8 -- Minimal Models for Chiral Active Matter

Benno Liebchen

Technische Universität Darmstadt, 64289 Darmstadt

**Status**

Chiral active particles (CAPs) are characterized by their ability to both self-propel and self-rotate, which results in spiralling trajectories [1,2]. In the absence of fluctuations, these trajectories take the form of circular (2D) or helical trajectories (3D), which are not superimposable on their mirror-image (hence the word "chiral"). CAPs are omnipresent both in nature and in the synthetic world: Many, if not most, motile microorganisms, from sperm cells to cilia, feature (slightly) asymmetric body shapes and hence, when self-propelling within a solvent, they are subject to hydrodynamic torques making them spiral. Similarly, shape-asymmetric synthetic active particles also self-rotate, as illustrated, e.g., by L-shaped swimmers. In addition, self-rotations can also emerge from spontaneous symmetry breaking as seen, e.g., in droplet swimmers [1,3].

When many CAPs come together, they feature a rich panorama of collective behaviours (see Fig. 1), which are currently under extensive exploration, particularly in the framework of generic minimal models describing CAPs with different types of interactions [1].

For isotropic short-ranged repulsions, CAPs can spontaneously self-organize into a dense liquid-like phase and a coexisting active gas (motility-induced phase separation (MIPS)), similarly as in straight swimmers. Here, the intrinsic particle rotations in CAPs tend to oppose MIPS in that the spinodal-like instability leading to MIPS shifts to higher self-propulsion velocities as the rotation frequency $\Omega$ increases. In addition, also new phenomena can occur in isotropically interacting CAPs that have no counterpart in straight swimmers [1], such as e.g. the emergence of finite sized dynamical clusters that counter-rotate with respect to the surrounding gas [4] and a chirality induced transition towards a hyper-uniform state [5].

CAPs with polar alignment interactions in turn form structures such as rotating macrodroplets (small $\Omega$) that select a characteristic system-size independent density (characteristic for phase separation), as well as microflock patterns (large $\Omega$) that select a characteristic length scale (characteristic for pattern formation) [6] (Fig. 1a). Other structures that emerge in aligning CAPs (with polar or nematic interactions) are polar vortices, "clouds" [7], and traveling waves [10] (Fig. 1b).

Mixtures of CAPs can feature additional phenomena such as species-segregation, or simultaneous phase separation in one species and pattern formation in the other one [8]. Beyond that, such mixtures serve as interesting model systems for synchronization. For instance, it has been found that active motion can significantly support or even induce synchronization as compared to, conventional non-moving oscillators and oscillators that move independently of their phase [1,9].

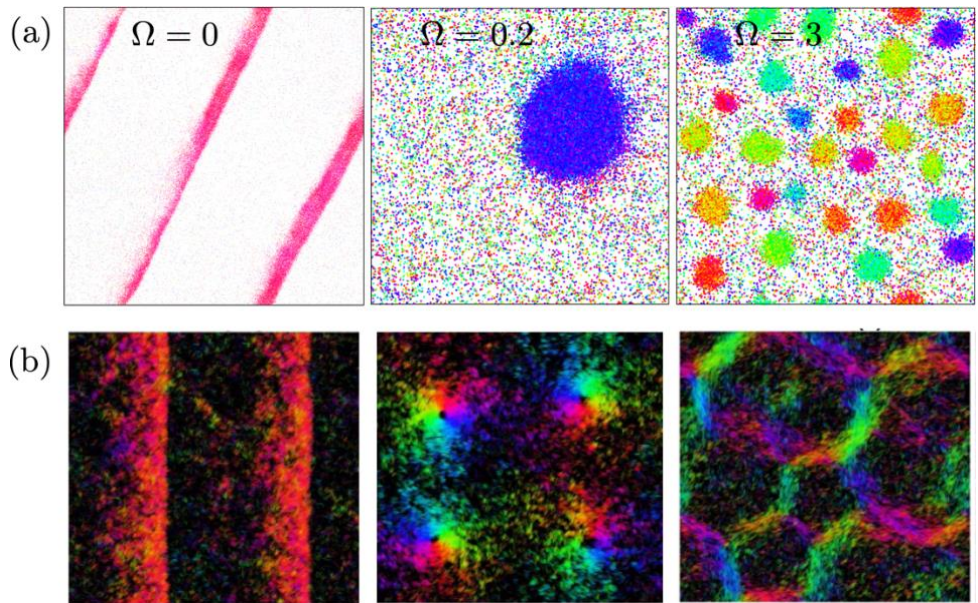

**Figure 1:** Simulation snapshots of CAPs with colors representing particle orientations: (a) Particles with polar alignment can self-organize into traveling bands (Ω = 0), rotating macrodroplets (Ω = 0.2) and micro-flock patterns (Ω = 3). (b) Particles with a symmetric distribution of angular frequencies can form (from left to right): traveling bands (surviving weak frequency dispersion), polar vortices (for polar alignment) and active foams (for nematic alignment). Figure from [1] which is adapted from [6,7].

**Current and Future Challenges**

Despite much progress in our understanding of the collective behaviour of CAPs, many important questions are still open [1].

A first challenge concerns the development of a precise understanding of the phase diagram of aligning CAPs. Here, the nature of the transition between the phase separated state, where the system selects a characteristic density (Fig. 1a, Ω=0.2) and the microflock pattern (Fig. 1a, Ω=2), where the system selects a characteristic length scale, is not yet well understood. In particle-based simulations, it has been observed that the length scale of the rotating microflocks slowly increases over time [6] which provokes the question if microflock patterns persist asymptotically for long times, or if they are ultimately converging to a macrophase separated state. In the latter case, an important question is if the mean cluster size increases according to the characteristic Lifschitz law, where the mean cluster size L increases with time as $L \sim t^{1/3}$, or if aligning CAPs show a different coarsening exponent (and perhaps also other unusual critical exponents). We are currently still lacking a comprehensive theory to systematically answer these questions.

For mixture of CAPs important questions regarding synchronization are still open. While particle-based simulations provide strong evidence that activity can support or even induce synchronization in systems with purely local interactions, we are currently lacking a comprehensive theory to rationalize active synchronization. One key question is if there is a proper synchronization (phase) transition or if active synchronization arises gradually.

Another challenge concerns the development of a systematic understanding of the nature of the limit from CAPs to spinners which only self-rotate but do not self-propel. A key question is: How does the collective behaviour of CAPs change when approaching the spinner limit? Related to this is the questions: While in spinners phenomena related to odd viscosity are currently intensively explored in experiments and theory, they have been hardly explored if circle and helical swimmers (with frictional contacts) albeit they would probably offer a rich phenomenology when accounting for frictional contacts between them.

Finally, a major open challenge concerns the link between effective "dry" models of CAPs and more microscopic descriptions that explicitly account for their self-propulsion mechanism and their full hydrodynamic (and phoretic) interactions. New coarse graining methods are required in the future to understand the net effect of hydrodynamic and phoretic cross-interactions in CAPs and to take them into account in increasingly realistic (minimal) models.

**Concluding Remarks**

On the theoretical side, much of what we know about the large-scale collective behaviour of chiral active matter, has been developed in the context of generic minimal models. Such minimal models will continue playing a key role in the future, both for particle-based simulations and for the development of increasingly comprehensive theories for chiral active matter.

Beyond the discussed open questions, many interesting generalizations of established minimal models are still lying ahead of us. For example, so far, most theoretical studies focus on CAPs with relatively simple interactions (isotropic short-ranged interactions or alignment interactions), whereas many experiments, from chiral Quincke rollers to chiral active colloids, involve more complex hydrodynamic and phoretic interactions that in general can be non-reciprocal, non-instantaneous and non-pairwise and could induce a new world of patterns.

Other generalizations of existing minimal models for the collective behaviour of CAPs could account for 3D effects, where CAPs follow helical trajectories, as well as for the influence of inertia, which could lead to different kinetic temperatures in coexisting phases. Finally, considering CAPs with frictional contacts could offer interesting links to odd viscosity related phenomena, that have so-far been primarily studied in spinners which self-rotate but do not self-propel.

# 9 – Active Elasticity & Active Solids

Jack Binysh[1] and Anton Souslov[2]

[1]Institute of Physics, Universiteit van Amsterdam, 1098 XH Amsterdam, Netherlands

[2]T.C.M. Group, Cavendish Laboratory, University of Cambridge, Cambridge CB3 0HE, United Kingdom

**Status**

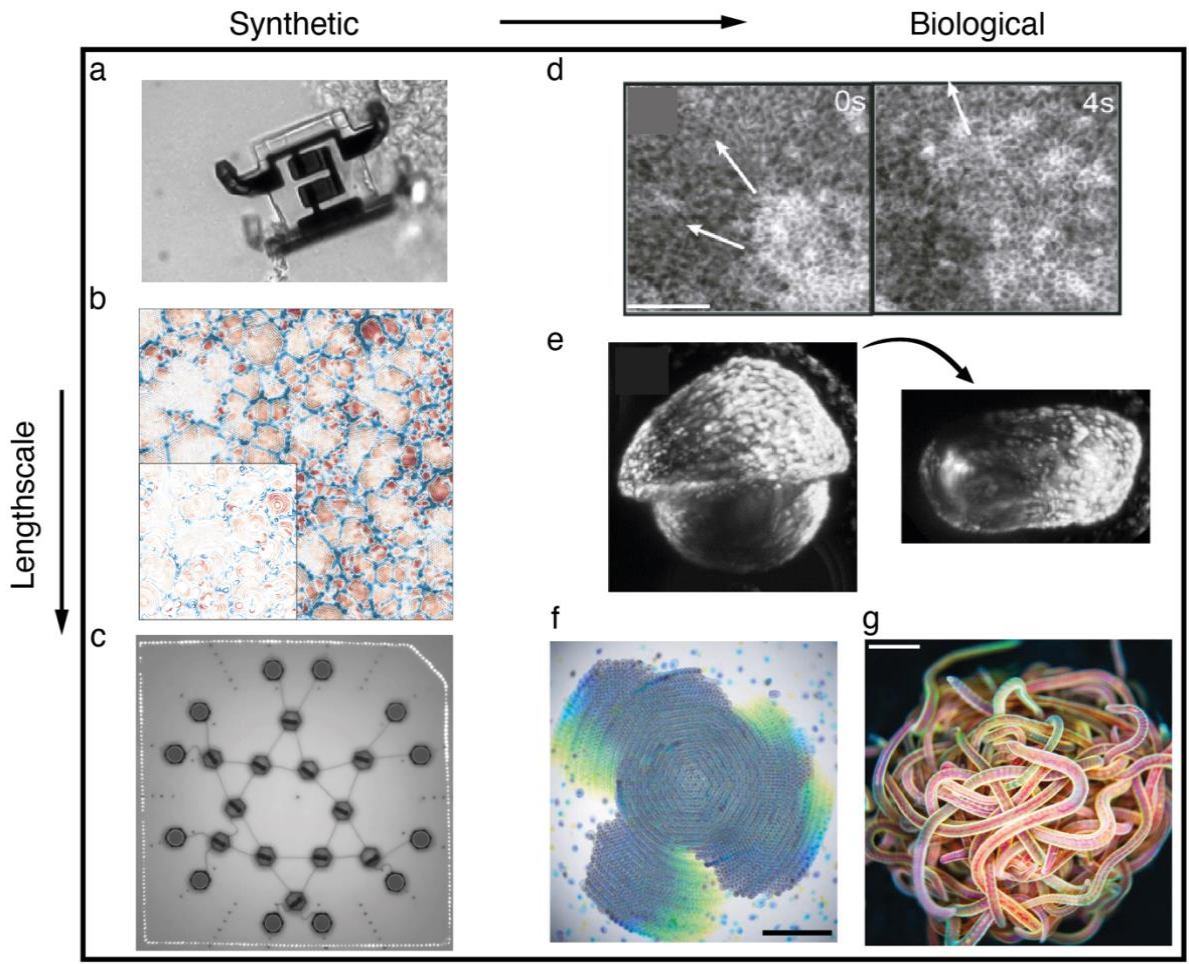

*Figure 5: Active solids exist on scales ranging from the micrometre to the macroscopic, and span synthetic (a-c) and biological (d-g) realizations. Systems composed of biological building blocks taken out of their natural environment (f) even blur the boundaries between the synthetic and the living. (a) Micrometre scale robots provide a route to miniaturizing insights from macroscopic active materials. Reproduced from [8]. (b) Activity melts a crystal composed of chiral spinners. Reproduced from [6]. (c) Robotic matter mechanically couples many robotic building blocks together, forming a distributed lattice capable of actuation and locomotion. Reproduced from [4]. (d) Epithelial cell layers act as active elastic sheets, pulsing with mechano-chemical contraction waves. Reproduced from [1]. (e) Volvox turns itself inside out via the mechanics of active thin sheets. Reproduced from [2]. (f) Starfish embryos cluster to form a living chiral crystal. The elastic excitations of this crystal propagate despite their overdamped nature, a feature powered by odd elastic moduli. Reproduced from [9]. (g) Worms cluster as a living, reconfigurable polymer melt. Reproduced from [3].*

Active solids are active matter with a reference state. This immediately places them in contrast to the paradigmatic examples of active matter, which are typically fluids. Here, we review how beautiful examples across living systems provide opportunity for fundamental physics and inspiration for engineered materials.

Within a single cell, a meshwork of polymers called the cytoskeleton is interlaced with molecular motors to form an active gel. This gel gives the cell mechanical rigidity but also acts as a dynamic, active material. Molecular motors slide polymers over one another, and these microscopic forces multiply up to macroscopic actuation of the mesh, allowing the cell to shape-morph and locomote. Zooming out from one cell to many, layers of cells stick together to form 2D elastic materials. But these sheets are not passive structures: they pulse with mechano-chemical contraction waves [**1**]. When a thin 2D sheet is embedded in 3D space, the sheet supports out-of-plane bending deformations as well as in-plane strains. This soft bending degree of freedom opens up a bewildering variety of shape formation across living matter. For example, the spherical micro-organism Volvox successfully turns itself inside out, a process orchestrated by the active layer of cells on its surface [**2**]. On the macroscale, examples of active biophysical solids range from army ants jamming as granular media to worm colonies clustering as living polymer melts [**3**].

The range of startling mechanical phenomena displayed by biophysical solids poses a clear physics and engineering challenge: how can we make artificial materials that act as machines? One design strategy mimics the creation of synthetic active fluids: to embed active elements into a passive solid [**4,5**]. This strategy has the advantage of drawing on the many developments in designing passive mechanical metamaterials. As an example of this approach, robotic matter is a type of active metamaterial that builds on the idea of swarm robotics. Many active units, for example, cm-scale robotic walkers called hexbugs, are elastically coupled to create a material somewhere between a solid and a robot. Freed from the constraints of equilibrium, the linear response of these active metamaterials features one-way wave amplification, and their elasticity exhibits anomalous odd moduli. Beyond the linear regime, active crystals display a rich variety of bifurcations and self-sustained nonlinear oscillations [**4,6**].

**Current and Future Challenges**

*From linear to nonlinear*
Classical fluid dynamics is rife with linear instabilities. The rich zoo of nonlinear dynamics exhibited by driven fluids provides vital context for understanding the instabilities of active fluids. By contrast, traditional solid mechanics focuses on linear response and oscillations about a steady state. The repertoire of familiar nonlinear states in passive solids is not broad enough to understand the instabilities of active mechanics. The exception that proves the rule here is the buckling of thin rods and sheets, and indeed many of the most notable examples of active solids are geometrically slender rods or shells [**2**]. A fundamental question is then: what nonlinear states arise in active solids due to the abundance of available instabilities?

*From self-propulsion to pulsation, rotation, and growth*
As well as classifying active solids by their instabilities, we can also classify by the form of the underlying activity. The examples of active solids above include activity through self-propulsion, but also encompass mechanical pulsation, persistent rotation, non-reciprocal couplings, growth, and ageing. Some examples involve a single active degree of freedom, others involve several coupled fields, for example, the coupling between rotation and propulsion in confined hexbugs [**4**]. In trying to classify the physical description of this variety of active systems, continuum approaches provide conceptual clarity: can we identify minimal continuum models which capture these behaviours?

*From the macroscale to the microscale*
The synthetic active solids shown in Fig. 1 are macroscopic systems, between mm and a few metres in size. By contrast, biological active solids span a range of lengthscales: the cytoskeleton is on the micrometre scale, whereas a growing redwood is hundreds of metres tall. To realise synthetic programmable materials, we need experimental platforms that miniaturize the dynamics of macroscopic active solids. This difference in lengthscale also entails a difference in physics: in a macroscopic active solid inertia is crucial, noise less so. By contrast, noise plays a crucial role in the dynamics of microscale active solids.

*From active to animate*
Living systems integrate actuation, sensing and computation throughout a single material. This triple of attributes, termed animacy, allows biological systems to precisely control the instabilities inherent in active materials and direct them towards purpose-driven behaviour. Achieving this level of control in synthetic systems opens the door to building materials that learn from their environment, reconfiguring their shape and dynamics to adapt to the task at hand. Synthetic active solids are well placed to deliver on the promise of animate materials, by weaving a layer of control into the reference state of the solid. However, explicit examples and strategies for controlling active solids remain a challenge.

**Advances in Science and Technology to Meet Challenges**

Fortunately, the design and fabrication of synthetic active solids can build on construction techniques for passive mechanical metamaterials. On the macroscale, combining 3D printing and readymade circuitry allows rapid prototyping of active solids on the mm to metre scales. Printing with electrically conductive polymers will dramatically expand the scope of this technique, allowing the design of active structures capable of large shape deformations. On the microscale, colloids coated in lipid bilayers can be selectively bound to one another to form floppy mechanical lattices, sheared by thermal fluctuations [**7**]. Driving such lattices via external fields or coloured noise opens one route to microscale active solids. However, what is truly remarkable about biological active solids is that they are emphatically *not* driven by some external field, or tethered to an external battery. Rather, they carry their own energetic reservoir with them, replenished when necessary via, for example, a supporting vasculature. One route to miniaturizing this combination of actuation and energy storage is using microscale robotics [**8**]. Another is by exploiting pre-existing biological engines to suit our own needs. For example, the crystal of starfish embryos shown in Fig. 1(f) supports overdamped waves, powered by each embryo's persistent rotation [**9**]. These waves may or may not serve a biological purpose, but we can treat them equally as a robust, purely physical phenomenon to be exploited.

In parallel to experimental advances, applications of control theory and machine-learning techniques [**10**] allow the forecasting and tuning of active fluid dynamics, as well as the solution of inverse problems in passive mechanical metamaterials. Combining these approaches to active solids appears a fruitful route to programming dynamical states of locomotion and actuation within these continuum materials.

**Concluding Remarks**

Active solids are a new frontier for fundamental and applied research geared towards designing new materials and understanding the mechanics of living matter. They motivate and stimulate work at the intersection of the life sciences, robotic design, dynamical systems, and mechanical metamaterials. Pushing the area forward requires a broad array of challenges to be surmounted, from theoretical questions of classification and synthesis to experimentally-driven advances in fabrication and prototyping. On a fundamental level, the presence of a reference state – to be deviated from, or reconfigured into – means that active solids combine the lively dynamics and flows of active fluids with a set of geometrical and topological intricacies arising from mechanics. What new phenomena will come from this interplay?

**Acknowledgements**
JB acknowledges funding from the European Union's Horizon Europe research and innovation programme under the Marie Sklodowska-Curie grant agreement No. 101106500. The authors acknowledge the support of the Engineering and Physical Sciences Research Council (EPSRC) through New Investigator Award no. EP/T000961/1.

# 10 -- Synthetic microswimmers with autonomous adaptation

Lucio Isa, Laboratory for Soft Materials and Interfaces, Department of Materials, ETH Zurich, 8093 Zurich, Switzerland

**Status**
Over the last two decades, synthetic microswimmers, and more specifically Janus microswimmers, have played a pivotal role in opening up completely new avenues of research in experimental active matter. In spite of their apparent simplicity, the complexity of their interactions, e.g., with boundaries and among particles, and the wealth of self-propulsion mechanisms have unveiled extremely rich physical behaviour. Part of the motivation to study synthetic microswimmers has also been their attraction as model systems able to capture essential features exhibited by their biological counterparts and to enable the realization of tantalizing opportunities in a broad field of applications, from smart delivery and transport, to autonomous sensing and remediation.

However, after all the wonderful progress to date, the field might have reached an appropriate moment to pose the question of whether the characteristics belonging to classical artificial microswimmers may no longer be just as simple as possible, but actually too simple, paraphrasing a famous Einstein quote. In particular, many of the steps that are associated to the far-reaching applications mentioned rely on the adaptation of the active units to a range of signals and stimuli as the key to enable switching and reconfiguring between different functions under different conditions. Remarkably, in living systems, adaptation happens in a fully autonomous way, in other words, relying only on internal feedback mechanisms and communication schemes. Nonetheless, directly taking inspiration from biology may not be the most efficient way to identify robust synthetic routes that lead to microswimmers with the desired characteristics. The diversity and complexity of biological signalling and feedback loops in fact complicates the efforts to distil minimal requirements to realize artificial microswimmers with autonomous adaptation. However, direct inspiration can be taken from another set of complex systems with autonomous, active and responsive units, where full control on their design and realization is already present: robotic systems. Minimalistic approaches have evolved towards the creation of "cyber-free" systems, possessing so-called "physical intelligence", where as much as possible of the actuation, sensing, computation, and response functionalities are carried out by the materials the robots are made from, without external logic and control [1].

**Current and Future Challenges**
While the concepts just introduced above are gaining increasing relevance as guidelines for the realization of macroscopic soft robotic systems, their translation to the colloidal scale remains highly demanding. In fact, even if tremendous progress has been achieved in miniaturization toward soft micro- or nano-robots, this nomenclature often considers objects where only one dimension is at the microscale.

In particular, two main challenges can be identified:

1) *The realization of reconfigurable artificial microswimmers with internal feedback schemes.*

A growing number of cases have emerged in the literature that show the collective reconfiguration of microswimmer assemblies, i.e., collections of active particles whose shape and motion can be externally modulated [2], but the reconfiguration of individual microswimmers is significantly less common. Rare examples include the reconfiguration of magnetic Janus cubes by toggling magnetic fields [3], bioinspired soft swimmers that reconfigure upon local changes of solute concentration [4] or colloidal clusters fabricated by sequential capillarity-assisted particle assembly (sCAPA) comprising soft, thermo-responsive microgels in programmable positions that change shape as a function of local temperature [5] (see Figure 1). The wide palette of different materials and the versatility of fabrication and assembly methods available for larger scale units cannot currently be found at the colloidal scale. Therefore, new design guidelines and approaches must be identified to translate macroscopic concepts such as movable joints, actuation via shape changes, detection of optical and chemical signals, etc., to the microscale. Moreover, while external feedback loops have been employed to regulate the dynamics of synthetic microswimmers [6], the integration of feedback schemes *within* the microswimmers themselves requires the use of responsive materials that can adapt their properties to given stimuli in a programmed fashion [5].

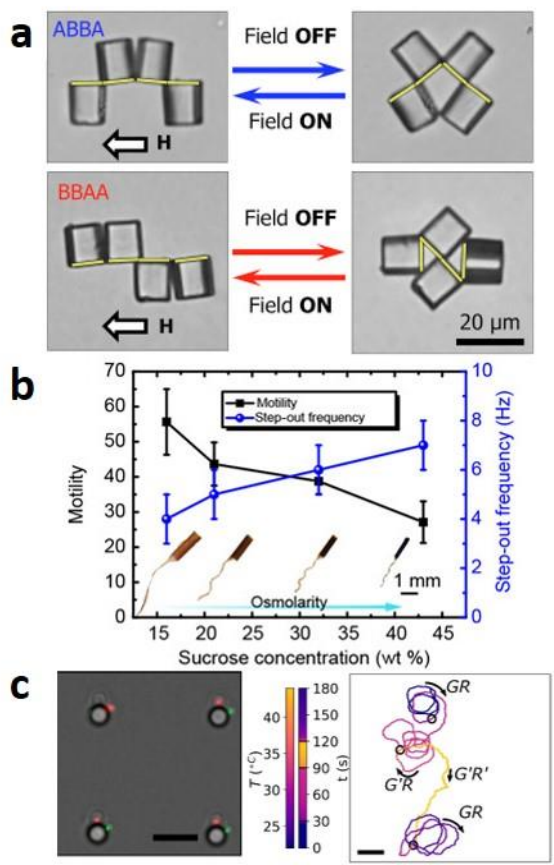

**Figure 6. a)** Reconfigurable microswimmers comprising magnetic Janus cubes. Adapted from [3]. **b)** Bioinspired magnetic soft microswimmers whose shape changes as function of sucrose concentration. Adapted from [4]. **c)** Reconfigurable colloidal clusters comprising temperature-responsive microgels whose swimming trajectory changes as a function of temperature. Adapted from [5].

2) *The implementation of artificial microswimmers with autonomous signalling schemes.*
Classical synthetic microswimmers interact via colloidal forces and hydrodynamics, and the most explored "communication" mechanism relies on chemotaxis, which is the capability of navigating a chemical gradient [7]. The inspiration comes again from biological entities, such as bacteria, which secrete signalling molecules and in turn trigger the chemotactic response of their neighbours to be attracted or repelled by it. Artificial chemotaxis, mainly studied from a theoretical perspective, has shown the spontaneous formation of a broad range of patterns. Albeit interesting, synthetic chemotactic signalling remains very basic and typically goes via local modulations of the propulsion "fuel" rather than from an orthogonal response and reconfiguration of the active particles. The open challenge is to develop new viable and programmable signalling pathways, where the transmission and reception of a signal is independent of the propulsion scheme.

**Advances in Science and Technology to Meet Challenges**
In order to meet the challenges highlighted above substantial progress is needed in the two following specific directions:

1) *Microswimmer synthesis and fabrication*

The realization of adaptive microswimmers with precise and programmable properties in large numbers via robust and versatile processes requires a paradigmatic shift of the way in which "colloidal engineering" has been carried out so far. Multi-material microswimmers are in fact necessary in order to integrate different functionalities and responses within single complex units. Classically, colloidal systems are realized via wet synthesis, modification and assembly methods. While the synthesis of a wide palette of colloidal particles with highly uniform and controlled shapes and compositions has become readily available, the integration of multiple materials within precisely controlled microstructures largely remains unattainable. However, classical strategies typically rely on self-assembly methods, which require fine-tuning of colloidal interactions and lead to poor final control or to very simple constructs. A very exciting way forward envisions combining colloidal self-assembly and directed assembly approaches with top-down nanofabrication methods to produce complex functional microstructures [8], e.g., as seen in Figure 2. Advancing in this direction requires the development of new responsive materials printable at the nanoscale and of new processes that enable the placement and integration of colloidal particles in three-dimensional microscale units with micro-nanoscale printed parts.

2) *Rethinking colloidal interactions*

The presence of feedback and communication schemes has profound implications on the description of systems of active colloids. In current frameworks, colloidal particles interact via reciprocal forces and instantly respond to the action of external potentials. Conversely, any feedback loop has a finite time response. If we think about the shape reconfiguration of a microswimmer or the detection of a chemical signal, these processes have an intrinsic time scale. Therefore, the adaptation of the microswimmer response to the spatio-temporal modulation of a corresponding signal will vary depending on its response rate [9]. New theoretical descriptions of the dynamics of microswimmers will have to include these aspects and new characterizations of response times will be required in order to inform the experimental design of microswimmers with targeted adaptive behaviour. Moreover, the integration of signalling schemes implies that the interactions between different microswimmers may no longer be reciprocal, in the sense that the behaviour of particle B may depend on particle A, but not vice versa. The theoretical exploration of non-reciprocal interactions has recently started and constitutes an exciting avenue for future research [10].

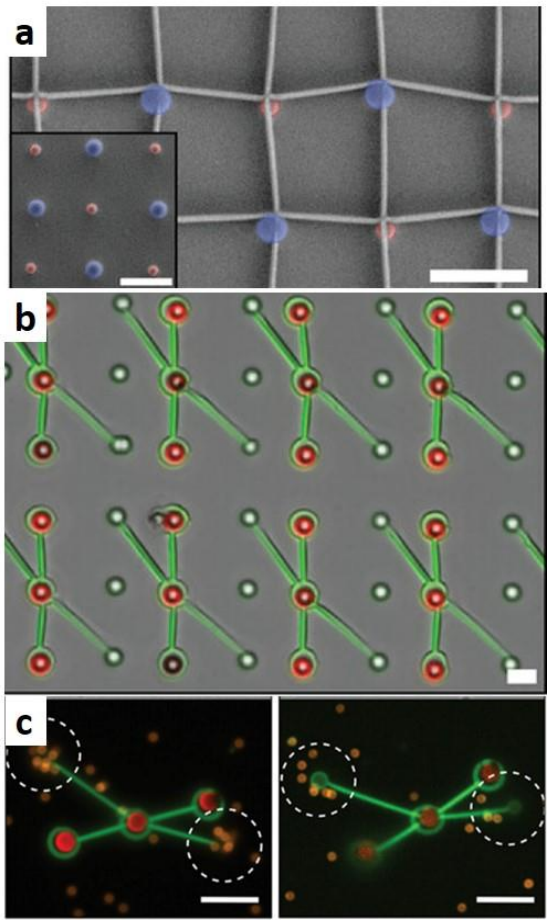

**Figure 7. a)** Example of multi-material microstrcutures with colloidal particles positioned in regular arrays by means of capillary assembly and connected by linkers printed by two-photon polymerization direct laser writing. **b)** Assembled and printed colloidal micromachines consisting of three 2.8 μm magnetic particles (red) and two 2.8 μm SiO2 particles (white) functionalized with DNA that are linked with printed photoresist (green). **c)** Example of selective capture and release of target particles (orange) as a function of temperature during "walking motion" of the colloidal micromachine. Adapted from [8].

**Concluding Remarks**

The realization of synthetic microswimmers with autonomous adaptation challenges the very core of how we synthesize and describe active colloidal systems. Not only it requires us to adapt, incorporate and translate technologies for micro and nano-fabrication to create a new class of microswimmers but also pushes the development and characterization of functional materials that can be integrated in such microswimmers. Here, the connection with robotic systems goes beyond the inspiration for minimalistic design approaches and forces us to include elements of feedback and control theory in new frameworks that describe the dynamics of those new active systems. Looking back, the introduction of the now classical Janus microswimmers two decades ago has generated tremendous progress in the physics, chemistry and materials science of active matter. With the emergence of new fabrication technologies, theoretical approaches and design guidelines, the field now finds itself at another turning point. Exciting possibilities for the translation of fundamental concepts in active matter to applications appear now closer in sight.

**Acknowledgements**
The author thanks Robert Style for discussions. This project has received funding from the European Research Council (ERC) under the the European Union's Horizon 2020 research and innovation program grant agreement No. 101001514.

**References**

*[(Separate from the two-page limit) Limit of 10 References. Please provide the full author list, and article title, for each reference to maintain style consistency in the combined roadmap article. Style should be consistent with all other contributions- use [IEEE style](IEEE style)]*

# 11 – Genetic engineering of active matter systems

Roberto Di Leonardo and Giacomo Frangipane, Dipartimento di Fisica, Sapienza Università di Roma

**Status**

Autonomous activity has been historically considered as a distinctive property of life. Today an increasing number of non-living, colloidal materials can generate systematic motion using energy that is stored internally or in the local environment. From the outside, synthetic self-propelled colloids appear remarkably similar to swimming bacteria and together have become the subject of a new field of physics that we call "active matter". Despite this external similarity, moving inside a cell, a static and homogeneous arrangement of atoms is replaced by a dynamic orchestra of thousands of biological machines executing a software program written in the DNA. After 50 years of extraordinary advances in genetic engineering we can now read that code, edit it, and write new code from scratch. A growing catalog of biological parts provides DNA sequences to implement sensors, actuators, signal transmitters and receivers, switches, and logic gates. These can be modularly recombined within model microorganisms, such as bacteria, to reprogram their motility response to environmental signals. In this respect, engineering new responses to light signals offers many advantages related to the ease of delivering light with high resolution in space and time. Using a natural light-driven proton pump, the proton motive force that powers the bacteria's flagellar motors can be controlled with light. Light patterns then translate into a space dependent motility that can be used for dynamic density shaping [1], optical rectification and confinement [2], transport of passive objects by light-induced active pressure gradients [3]. In addition to using structured light environments to guide bacterial pattern formation, one can also reverse the logic and decode bacterial patterns to recover environmental factors. *Proteus mirabilis* was engineered to modulate the expression of swarm-related genes based on one or more environmental chemical signals. The resulting patterns have been shown to contain a record of environmental history [4]. The vast protein repertoire of cells makes it possible to extend control to many other mechanical responses besides motility. Using optogenetic tools, these responses can be then tuned in time and space. By placing a key metabolic enzyme under optogenetic control, cell growth rate can be regulated over time [5]. Bacterial biofilms can be sculpted by projecting an optical template over cells expressing adhesin proteins under light [6]. But the domain of gene editing can go beyond single-cell responses to engineer interactions and program the spontaneous generation of multicellular patterns. Pairs of complementary synthetic adhesins have been used to implement cell-cell adhesion logic in swarming bacteria and to program multicellular interface patterns [7]. By detecting chemical signals emitted by other cells, bacteria can regulate gene expression in response to cell population density. Exploiting the quorum sensing system from *Vibrio fischeri* to couple density and motility, *Escherichia coli* cells have been programmed to form periodic stripes of high and low cell densities [8]. This approach has recently been extended to a multicomponent bacterial system with reciprocal regulation of motility, resulting in programmable self-segregation or co-localization in periodic stripes [9].

**Current and Future Challenges**

**Expand the protein pool.** Expanding and optimizing the protein pool for motility control in microorganisms is certainly a major challenge. For example, a big constraint in utilizing proteorhodopsin for speed control is the necessity for an oxygen-deprived environment. Moreover, speed control certainly does not exhaust the possibilities of optical motility control. For example, the ability to modulate tumbling rate would provide direct access to the chemotaxis system and allow it to be reprogrammed by light. Collective motility states, such as those found in active turbulence or swarms, have been described by coarse-grained theories whose parameters are often only phenomenological. The possibility of precise parametric control of characteristics such as velocity, tumbling rate, and density can facilitate linking these macroscopic parameters to the microscopic characteristics of the individual microswimmers.

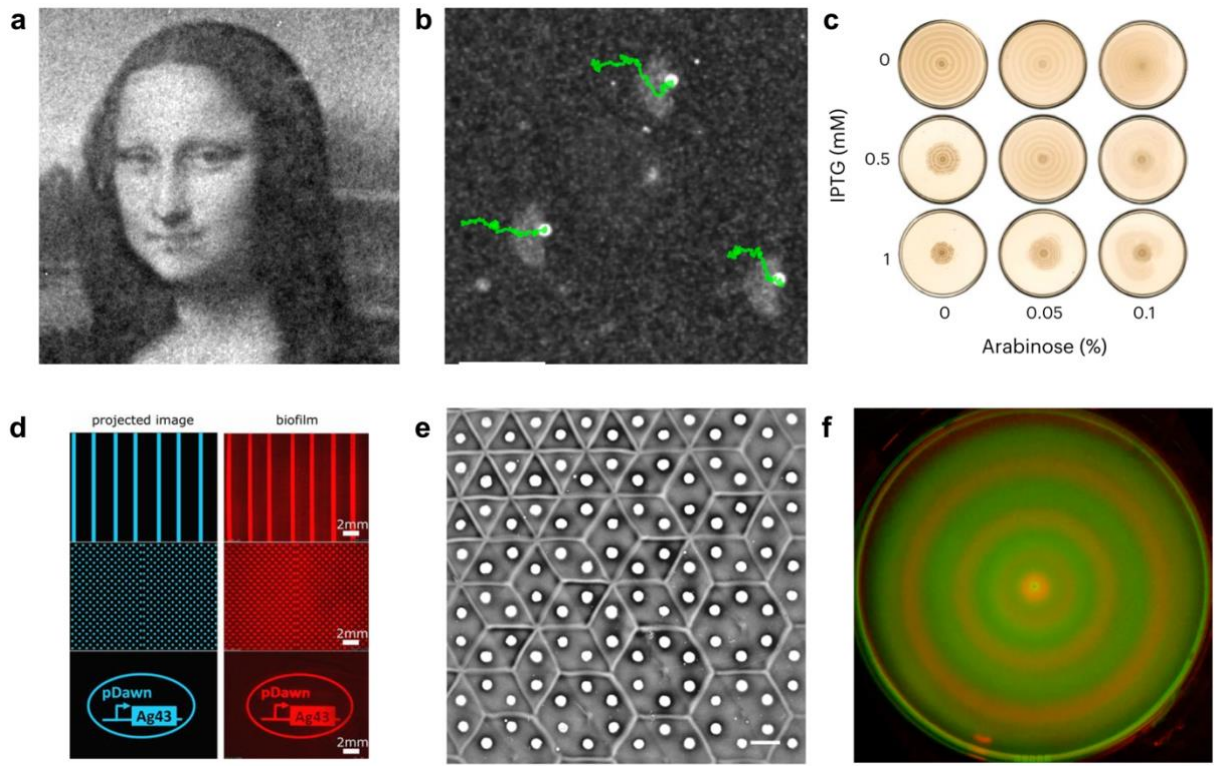

**Figure 1. a)** Speed control in engineered bacteria can be used to shape density (adapted from [1]) or **b)** generate gradients of active pressure (adapted from [3]). **c)** Engineering swarm patterns as spatial records of environmental inputs (adapted from [4]). **d)** Optical lithography of bacterial biofilms (adapted from [6]. **e)** Programmed multicellular interface patterns (adapted from [7]). **f)** Self-segregation of engineered bacteria strains with reciprocal motility regulation (adapted form [9]).

**Integrate multiple signals.** As the space of controllable responses will increase there is the need to integrate multiple control signals. On the one hand, it is necessary to have decoupled sensors, on the other hand, the recorded signals need to be integrated through an internal logic that can implement the desired response.

**Expand the microorganisms pool.** Model organisms such as *E. coli* are relatively easy to engineer using the streamlined tools for gene editing. However, there is a huge variety of microorganisms that have adapted to all possible environmental conditions. Pathogens can often recognize and target specific niches or host tissues within the human body during infection. Magnetotactic bacteria have a built-in compass, while microalgae can "see" the direction of light. Some bacteria can rapidly expand by a cooperative motile behavior called swarming. Understanding and reprogramming their motility could have crucial applications for active matter, with applications in many other scientific fields, such as micro-robotics and engineered bacteria therapeutics.

**Engineer microorganisms consortia.** Different bacterial strains can be engineered to have varied functionalities. By combining these strains, a consortium can exhibit a wider range of behaviors and responses to environmental stimuli than a single strain. Through the interaction of multiple motility-regulated strains, consortia can self-organize into intricate structures, which is valuable for understanding pattern formation in biological systems and for potential applications in biomaterials.

**Advances in Science and Technology to Meet Challenges**

Advances in software tools for protein design will be crucial to explore a broader space of sensor and transducer proteins in order to expand the possibilities of controlling motility and interactions. New control devices can be obtained by chimeric constructs that combine receptors and transducer domains from different proteins across different species. A fusion chimera combining an archaeal photoreceptor with a bacterial chemotaxis protein has been shown to generate phototaxis responses in *Escherichia coli*. Synthetic biology has demonstrated the ability to exploit the computational capability of gene regulatory networks to program spatial patterns of gene expression in response to one or more external signals. A promising avenue for future research lies in the opportunity to harness the computational power of cells for the development of more sophisticated and responsive dynamic behaviors in active matter. Thanks to recent advances in optogenetics, bacteria can distinguish between red, green and blue light and respond by modifying gene expression [10]. Using color light signals, researchers could orchestrate a precise program of cellular behaviors such as spreading, adhesion, growth, and release of synthetic biomolecules. In this regard, advances in cybergenetics, an emerging field of research combining control engineering and synthetic biology, will be instrumental in introducing novel platforms for optical control, along with the development of new fundamental principles and algorithms. Optogenetic tools, however, have inherently slow response times, related to gene expression, ranging from several minutes to hours. To achieve faster response times, existing proteins can be modified by inserting a light-sensitive domain that can be used to rapidly turn their activity on and off. This could be a promising route for quick and space dependent switching between different modes of motility. Expanding the pool of engineered microorganisms could enrich the set of "smart atoms" to build new active materials, but it requires the development of new tools for genetic engineering.  Although for some microorganisms, such as microalgae, these tools are emerging at a faster rate today, they are still underutilized. Finally as new engineered microorganisms enter the domain of biological active matter, new possibilities will arise for distributing tasks among different species by programming new communication mechanisms. As an example, a leading species could detect environmental signals and convert them into cues for a secondary species to engage in an assisted form of taxis.

**Concluding Remarks**

Much of the research on active matter delas with systems of self-propelled particles that interact with themselves and the environment through mechanical and hydrodynamic forces. These simple ingredients are sufficient to generate a broad spectrum of dynamic behavior that is unmatched in systems at equilibrium and which, for this reason, are often labeled as 'living'. Today we know that even simple living entities, such as bacteria, are not just self-propelled 'monads' that blindly collide with one another. They constitute a grid of computing machines, exchanging information in the form of chemical signals and responding to environmental challenges with a coordinated behavior far more complex than swirls and flocks. On the one hand we have active matter physics researchers putting together the pieces of a new, out-of-equilibrium dynamic that often neglects internal degrees of freedom. On the other hand, synthetic biology emphasizes the 'biochemistry' of these internal degrees of freedom, often modeling spatial interactions as ordinary diffusive processes for both cells and molecules. It is only through the convergence of these two endeavors that active matter will fulfill its promise and develop into a quantitative science of 'living' systems, a science that is able to explain and design complex behavior resulting from the interaction of external and internal degrees of freedom.

**Acknowledgements**
R.D.L. acknowledges funding from European Research Council under the ERC Grant Agreement No. 834615.

# 12 -- Towards clinical applications of magnetic micro/nanorobots

Hongri Gu[1] and Bradley J. Nelson[2]

[1]Department of Physics, University of Konstanz, Konstanz, Germany
[2]Institute of Robotics and Intelligent Systems, ETH Zürich, Zurich, Switzerland

**Status**

Over the past decade, magnetic micro- and nanorobots have gradually made their way from science fiction to potentially powerful medical robotic systems in clinical applications[1]. There has been a noticeable shift in the research community's focus from curiosity-driven questions, such as how to move these tiny robots[2], to the challenges of specific real-world applications, such as how to harness the benefits of microrobots for the treatment of specific diseases and integrate them into the surgical workflow[3–5]. With intensive cross-disciplinary collaboration, the path to potential medical applications in a realistic clinical setting is becoming clearer[6]. Here we discuss the three critical components in a microrobotic system and research focus shifts we see based on our recent observations.

First, we consider the micro/nanorobots themselves. Given their microscopic size, a single microrobot is often not sufficient for tasks such as drug delivery; instead, a large number of microrobots in the form of swarms are necessary[7]. This requirement has led to a shift in research focus from a single delicate microrobot that is fabricated, for example by Nanoscribe, to mass production methods and scalable control strategies for large swarms. Another shift in research focus is on the material. In the early days of magnetic microrobots, nickel and cobalt were often used because of their compatibility with existing microfabrication processes. However, these materials cannot safely degrade in the human body. The research community is gradually moving towards biodegradable magnetic materials such as FePt and iron oxide nanoparticles.

The second component is the magnetic actuation system, which typically consists of a few permanent or electromagnets. This aspect is overlooked in most papers, where the reader's attention is drawn to the mesmerizing "dancing" microrobots (and we cannot blame them for this!). Many small home-built magnetic actuation systems are sufficient for on-chip demonstrations, but none of them function at human clinical dimensions. Scaling up the workspace of a magnetic actuation system requires scaling up the surrounding magnets, whose size, weight, and cost increase rapidly. For example, the Aeon Phocus (maximum magnetic field: 130 mT) weighs more than 7 tons, which requires a special installation in the hospital. Therefore, the focus has shifted to finding solutions for a portable and powerful electromagnetic system. A recently available magnetic actuation system, Navion from Nanoflex Robotics, uses patented technology to increase the power density in electromagnets. The portable system balances magnetic field strength, frequency, and workspace while being small and light enough to be easily moved in and out of the operating room for a wide range of applications.

Finally, we must consider the imaging system that tracks the microrobots inside the body in real time. A C-arm X-ray fluoroscope is often chosen for this imaging purpose, but its resolution is limited to the millimeter scale. Magnetic resonance imaging systems can provide higher imaging resolution. However, the strong magnetic materials on the microrobots pose risks and distort the image[4]. These imaging systems must also make trade-offs between spatial and temporal resolution, and controlling the motion of microrobots often simultaneously requires high spatial and temporal resolutions. A shift in research focus is to develop new non-invasive imaging techniques or to combine multiple imaging modalities to improve the overall results. Some techniques (magnetic particle imaging, photoacoustic imaging, etc.) have emerged in the last decade and show great promise. Unfortunately, there are no human-sized systems available on the market yet.

**Current and Future Challenges**

Based on these three microrobot challenges, i.e., fabrication, actuation, and sensing, we can consider how to design the material composition of magnetic micro/nanorobots themselves. As the core part of the system, magnetic micro/nanorobots must have at least three types of materials on board: (1) Magnetic materials: These materials are crucial as they provide the necessary magnetic force and torque required to efficiently navigate through bodily fluids and reach their intended destination. (2) X-ray contrast agents: Integrating X-ray contrast agents into the micro/nanorobots is essential to enable their visualization through fluoroscopic imaging, which aids in tracking and monitoring their movement within the body. (3) Drug and carrier materials: These materials serve as a structural framework and reservoir for functional drug components. They ensure that the drugs are held securely during transport and are released precisely at the target site when needed.

Given the tiny size of microrobots, a tradeoff must be made for the limited on-board space among the three essential types of materials. Increasing the amount of magnetic materials on board leads to improved locomotion speed and controllability in complex in vivo environments. A higher concentration of X-ray contrast agents improves tracking capabilities while reducing radiation exposure to patients. Increasing the drug load can increase the local effective drug concentration, potentially improving treatment efficacy.

This situation poses a significant challenge in microrobot design, creating a dilemma in which improving one aspect of performance inevitably comes at the expense of others, necessitating compromises. Some microrobots may excel in certain aspects while falling short in others, making it difficult to strike the right balance. For example, hydrogel-based microrobots exhibit excellent biodegradability and drug delivery capacity, but their limited magnetic material content (surface-coated $Fe_3O_4$) hinders their ability to overcome in vivo fluid flow[8]. Another example is the use of pure nickel, which has tremendous propulsive power enabling them to swim against blood flow. However, drugs can only be bound to the surface, and the microrobots cannot degrade in the body[9].

Addressing the challenge of microrobot design and multi-object optimization is a task that requires out-of-the-box thinking and innovative approaches. A recent example is a structured fiber with embedded micromagnets, called artificial microtubules, which provides robust and guided transport of a swarm of microrobots[10]. This approach allows the microrobot to achieve significantly higher swimming speeds while maintaining the same driving frequency and magnetic field strength. Importantly, this speed increase is not achieved by increasing the amount of magnetic material or by developing a larger and stronger external magnetic actuation system, which could be cost prohibitive. Instead, the improved speed results from the introduction of novel magnetic interactions between an embedded micromagnet array and the mobile microrobots.

**Concluding remarks**

In this rapidly evolving field, the compelling potential of micro/nanorobots is attracting growing interest from physicists, engineers, medical doctors, and investors. As researchers continue to push the boundaries and achieve breakthroughs in advancing the clinical applications of these tiny engineering marvels, the prospect of revolutionizing the healthcare sector looms ever larger. As part of this community, we are excited about recent breakthroughs in the field and the long-awaited realization of in-human applications.

**Acknowledgements**

B.J.N. is a co-founder of Nanoflex Robotics AG and MagnetbotiX AG. B.J.N. is an inventor on patents related to microrobots and magnetic navigation systems. H.G would like to acknowledge the support from Swiss National Science Foundation (project number: 203203).

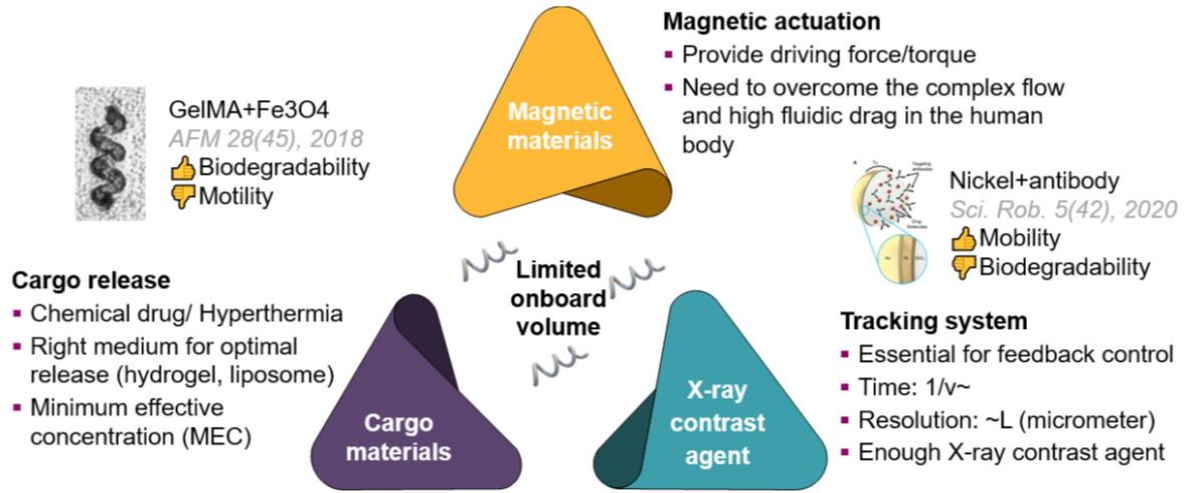

**Figure 1. A dilemma in microrobotic design.** For magnetic microrobots, three types of materials are essential for in vivo applications. Adding more materials to improve one aspect will inevitably compromise performance in another aspect.

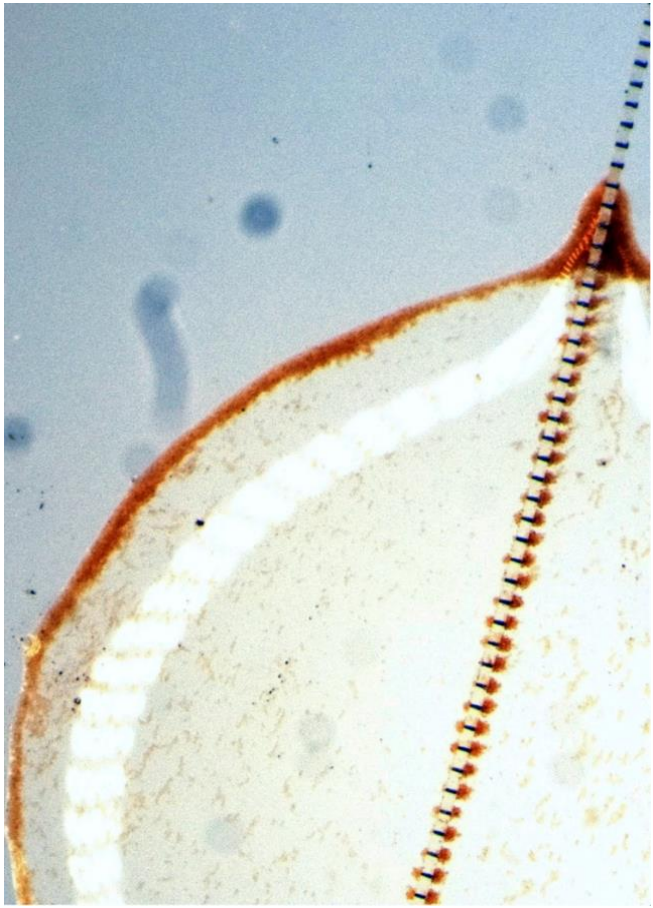

**Figure 2.** Collective transport of magnetic microparticles on an artificial microtubule near a droplet.

# 13 – Non-reciprocal interactions in active matter

Fridtjof Brauns[1] and M. Cristina Marchetti[2],

[1]Kavli Institute for Theoretical Physics, University of California Santa Barbara, California 93106, USA

[2]Department of Physics, University of California Santa Barbara, California 93106, USA

**Status**

Newton's third law establishes that pair interactions are reciprocal: for every action there is an equal and opposite reaction. The fundamental forces of nature, such as the gravitational force, obey Newton's third law. In living systems, however, interactions are often nonreciprocal due, for instance, to the asymmetry of sensory perception or information transmission [1]. Nonreciprocity is also ubiquitous in active, open and nonequilibrium multi-component systems, where it can result in a wealth of new phenomena that are currently the subject of intense investigation [1-4]. At the microscopic level, nonreciprocity corresponds to the breaking of detailed balance. At the mesoscopic level it results in couplings that cannot be obtained as derivatives of a Hamiltonian or free energy. We specifically distinguish two types of non-reciprocity that have been considered in the literature: (i) in the pairwise coupling between agents/particles in a "single species" system (for instance in mechanically coupled actuators or flocking models with vision-cone-based interactions), (ii) in the cross-coupling between multiple species/fields (see Fig. 1). We mostly focus on the latter case here, although there is also considerable interest in the former.

Recently, attention has focused on the role of nonreciprocal interactions in extended many-body systems with hydrodynamic modes due to conservation laws and symmetries (Goldstone modes), where nonreciprocity drives temporals organization, setting static patterns in motion and resulting in the emergence of a rich class of new dynamical steady states. These systems identify a new class of pattern formation [5] and offer the opportunity for unifying apparently different realizations, such as active/passive particle mixtures, active gels and reaction–diffusion systems [2,6].

The common mechanism underlying the emergence of dynamical patterns from the interplay of nonreciprocity and hydrodynamic modes can be highlighted in the context of diffusion of conserved concentration fields, where fluctuations of two diffusively coupled conserved densities can only decay through redistribution of the globally conserved masses. Remarkably, antagonistic cross diffusivities of conserved fields yield a chase-and-run behaviour, resulting in a dynamical steady state of directed traveling density waves [2,3,6] (see Fig. 1, left). The phenomenon is similar to that observed in

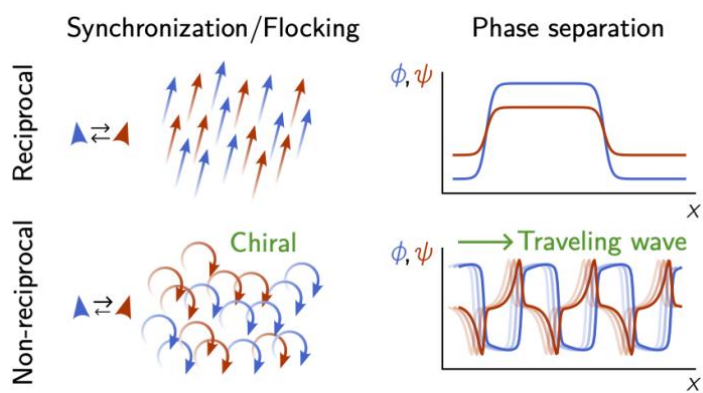

**Figure 1.** The interplay of non-reciprocal coupling with hydrodynamic modes due to Goldstone modes (rotation invariance in synchronization/flocking, left) or conservation laws (in phase separation, right) can lead to dynamic phases such as chiral rotations (left) [4] and traveling waves (right) [2,3,6]. antagonistically coupled flocking agents, where birds A want to align with birds B, but B wants to anti-align with A. In this case the frustration arising from the nonreciprocal interaction restores the broken polar symmetry of the flocking state, organizing a chiral state where the birds chase each other tails [4] (see Fig. 1, right). Both phenomena are identified by exceptional points in the dispersion relation of the linear fluctuation spectrum [4] and are in fact mathematically equivalent on the level of a two-mode approximation [2]. However, beyond this elementary setting, conservation laws give rise to striking new phenomena [6].

**Current and Future Challenges**
Recent work has proposed the non-reciprocal Cahn-Hilliard equation (NRCH) as a model of spatio-temporal order in systems with conservation laws [2,3]. The Cahn-Hilliard equation provides a mean-field description of phase separation in terms of the dynamics of a conserved volume fraction or density field. It captures the kinetics of spinodal decomposition and coarsening as well as bulk phase separation in the long-time steady state. The minimal NRCH model is obtained by coupling the phase separating Cahn-Hilliard field to a second diffusing field with nonreciprocal cross diffusivities. It has been shown to provide a description of a number of physical systems, including mixtures of active and passive Brownian particles, mass-conserving reaction-diffusion systems, such as the Min-protein system of *E. coli*, and viscoelastic gels driven by isotropic active stresses [6]. Even in its simplest realization, it exhibits a remarkable array of new dynamical states, from traveling density waves with unusual coarsening behaviour to undulating fronts and oscillating structures (see Fig. 2).

Many open questions remain: Only deterministic incarnations of the NRCH and related models have been considered so far, and found to exhibit multistability. An important open question is understanding the role of noise, especially as a possible mechanism for pattern selection, as known to be the case in certain classical flocking models [7]. The NRCH model contains a single length scale—the interfacial width. Yet it exhibits arrested coarsening, with a new mechanism for nonlinear pattern selection yet to be fully understood. More generally, while we have a systematic classification of pattern forming systems according to symmetries and conservation laws [8], we lack general principles for a similar classification of dynamical time-ordered states. Attempts based on bifurcation theory and normal forms [5] are useful, but the number of parameters increases rapidly with the order (co-dimension) of the bifurcation, making systematic exploration difficult. Whether more practical, physics-informed classifications are possible is an open question.

Non-reciprocity provides a new way to drive systems away from equilibrium, which is distinct from the classical active stresses considered in active polar and nematic systems. This has been exploited for instance in the case of active solids and metamaterials, but is largely unexplored in active fluids, where it may lead to new classes of active polar and nematic liquid crystals.

Finally, an open challenge is understanding how nonreciprocity arises upon coarse-graining. In this context it is natural to ask whether there are common features between nonreciprocal interactions within one species (i) or among species (ii), and whether (ii) might emerge on a coarse-grained level from (i).

**Advances in Science and Technology to Meet Challenges**
Theory and computation have advanced beyond experiments. An important challenge going forward will be developing experimental platforms that realize and test theoretically predicted phenomena in non-reciprocal systems. Non-reciprocal mechanical systems have been developed to study phenomena like odd elasticity and edge modes [9]. However, the size elementary mechanical units in these systems limits current experiments to small systems with 10s or 100s of units [10]. Overcoming this limitation will be key to study collective phenomena on large scales and make contact with hydrodynamic theories. An alternative to engineering macroscopic systems is to employ living matter such as bacterial suspensions, reconstituted filament-motor mixtures or protein-based reaction–diffusion systems. The *in vitro* Min system is a particularly promising candidate for the latter. Advances in biochemical protein and lipid design will enable increasingly fine-grained control over the molecular interactions which will allow testing key predictions from the NRCH model.

Tailored nonreciprocal interactions may provide a strategy for controlling the direction of molecular interaction and resulting assembly to create materials capable of transitioning between desired target states. Multicomponent active colloidal suspensions are a promising venue for engineering nonreciprocal interactions and examining their implications on self-assembly.

Finally, recent research is beginning to suggest a connection between nonreciprocal active matter and open quantum systems. In quantum settings nonreciprocity can arise from the interplay of dissipation and driving in open systems, as well as in quantum systems subject to local measurements. In both active matter and quantum many-body systems nonreciprocity can yield sensitive dependence of the response to external perturbations and provide a key role in information processing and sensing.

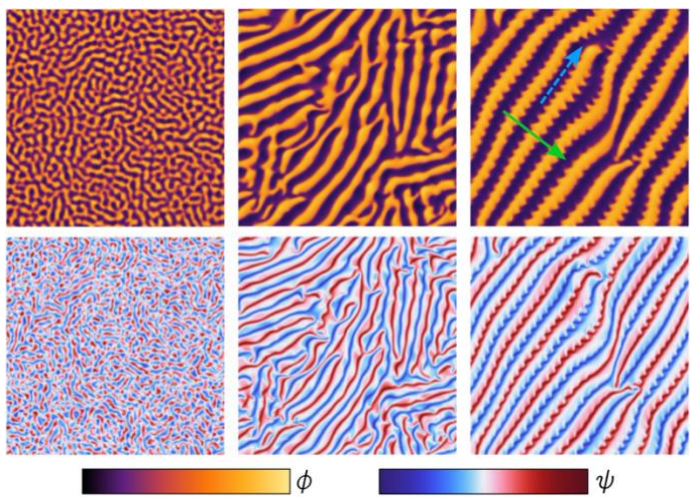

**Figure 2.** Formation of undulating traveling waves in the non-reciprocal Cahn–Hilliard model [6]. The undulations propagate along the wavefronts (blue dashed arrow), transversally to the direction of wave propagation (green arrow).

**Concluding Remarks**
Studying non-equilibrium systems from the point of view of non-reciprocal interactions has enabled a unified understanding of previously studied systems and explained the common principles underlying the emergence of dynamic phases such as traveling waves and chiral flocking. Systematic theoretical and computational studies of minimal models of non-reciprocally coupled systems with Goldstone modes and conservation laws have revealed a wealth of new phenomena that we are only beginning to understand. While experimental realizations are still limited, several platforms, including mechanical and bio-chemical systems, will allow to test theoretical predictions and may make it possible to utilize non-reciprocity as a design principle to engineer new functional materials.

**Acknowledgements**

FB was supported by Simons Foundation grant (#216179) to the KITP. MCM was supported by the National Science Foundation award No. DMR-2041459.

# 14 – Pursuit and Swarming of Cognitive, Self-Steering, Active Particles

Gerhard Gompper, Theoretical Physics of Living Matter, Institute for Advanced Simulation, Forschungszentrum Jülich, 52425 Jülich, Germany

**Status**

Many biological organisms, from uni- and multi-cellular micro-organisms to insects and larger animals, have the unique ability of directional sensing of their environment, the cognitive processing of this information, and the subsequent adaptation of their motion. Examples on the cellular level are neutrophils (white blood cells) hunting bacteria in blood and sperm cells homing in on the egg; on macroscopic scales, it is the flocking of birds and schooling of fish. The design of artificial agents with similar capabilities is also important for the development of microrobotic systems [1].

From the modelling side, the investigation of principles and mechanisms of swarming has started with the suggestion of the "boid" model by Reynolds [2], the "alignment" model by Vicsek [3], and the "zonal" model by Couzin [4]. The main idea in all these models is that agents adapt their direction of motion according to information input about location and orientation of their neighbors. Here, the Vicsek model focusses on alignment interactions, while the other models also consider reorientation and motion toward the center of a swarm to achieve swarm cohesion [2,4,5].

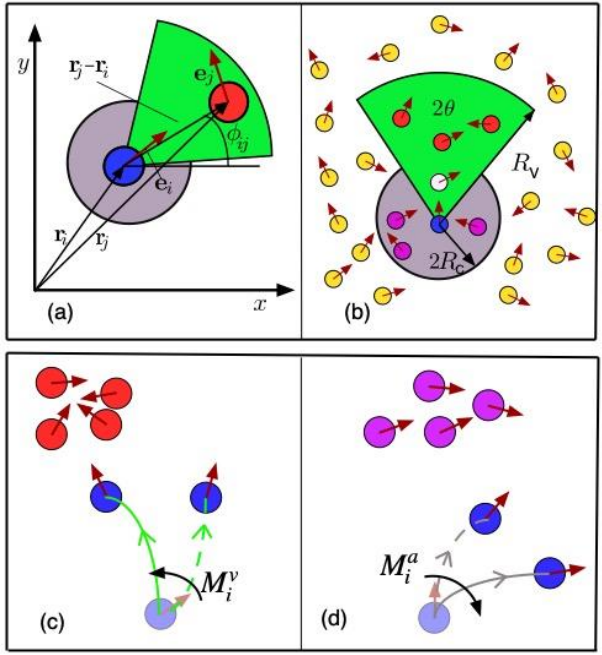

Figure 1: (a) Schematic representation of vision cone and alignment neighborhood of particle *i* (blue) with orientation $e_i$, distance vector $r_{ji}=r_j - r_i$ to other particles. (b) Polar orientation field (grey) with cutoff $R_C$, and vision cone (green) with vision angle θ and vision range $R_V$. (c) Steering of an iABP (blue) by adaptive vision-induced torque $M_i^v$, for weak and strong maneuverability $\Omega_V$, shown by dashed and full green trajectories, respectively (d) Steering by alignment-induced torque $M_i^a$, for weak and strong maneuverability $\Omega_a$, with dashed and full grey trajectories, respectively. Adapted from Ref. [7] with permission.

More recently, we have considered a minimalistic model for the collective behavior of self-steering, self-propelled agents (intelligent active Brownian particles, iABPs), which emphasises the physical aspects of swarming [6,7]. iABPs move with constant speed $v_0$ and propulsion direction **e**, where the temporal evolution of the latter is determined by $d/dt\, \mathbf{e}_i = \Omega_v \mathbf{M}_i^v + \Omega_a \mathbf{M}_i^a + \Lambda_i(t) \times \mathbf{e}_i$, with the vision-induced and alignment-induced steering torques $\mathbf{M}_i^v = N_{c,i}^{-1} \sum_j \mathbf{e}_j \times (\mathbf{u}_{ij} \times \mathbf{e}_i)$ and $\mathbf{M}_i^a = N_{a,i} \sum_j \mathbf{e}_j \times (\mathbf{e}_j \times \mathbf{e}_i)$, respectively, as illustrated in Fig. 1. Here, $\mathbf{u}_{ij} = \mathbf{r}_{ij}/r_{ij}$ is the unit vector from particle *i* to particle *j*. The first term on the right-hand side describes the vision-induced steering torque toward the location of other particles in the vision cone of particle *i* with maneuverability $\Omega_v$, while the second term describes the steering torque for alignment of the particle orientation with those of neighboring particles in the alignment range with maneuverability $\Omega_a$. The last term describes orientational noise. The steering toques are normalized by the numbers of particles $N_{c,l}$ and $N_{a,l}$ in the vision cone and alignment range, respectively. This implies that the interactions are *non-additive*, and the vision-induced steering is also *non-reciprocal*!

It is important to notice that even such a minimalistic model of directional sensing and self-steering contains a significant number of parameters: the propulsion speed $v_0$ of individual agents, characterized by the Peclet number $Pe=v_0/(\sigma D_r)$ with particle diameter $\sigma$ and rotational diffusion coefficient $D_r$, the vision angle $\theta$ and vision range $R_v$, the maximum torques allowed for varying the propulsion orientation due to vision-induced steering and alignment, characterized by the maneuverabilites $\Omega_v$ and $\Omega_a$, respectively, and the particle volume fraction $\phi$. Visual perception is meant here as a stand-in for any kind of directional sensing, from real vision to chemotactic sensing.

Examples of simulation results [7] for this model are shown in Fig. 2. The tendency for the formation of compact aggregates, elongated worm-like swarms, and millings, is strongly affected by the vision angle $\theta$ and the maneuverability ratio $\Omega_a/\Omega_v$. Milling structures have been observed in many animal groups, such as fire ants, reindeers, and fish.

The same type of model can also be employed to study the pursuit of a target (or prey) by a noisy pursuer [8,9]. In this case, the effect of hydrodynamic interactions has also been investigated, where the active particles are described as squirmers [9]. Two interesting examples of the role of hydrodynamics are [9]: (i) For two squirmers with vision-induced steering, a faster pusher-pursuer may not be able to reach a slower pusher-target due to hydrodynamic repulsion; a similar effect occurs for fish larvae unable catch the food -- known as "hydrodynamic starvation". (ii) For two squirmers with alignment steering, speed adaptation of the pursuer is required to form long-lived squirmer doublets; in this case, the pursuer can "steer" the target through hydrodynamic torques.

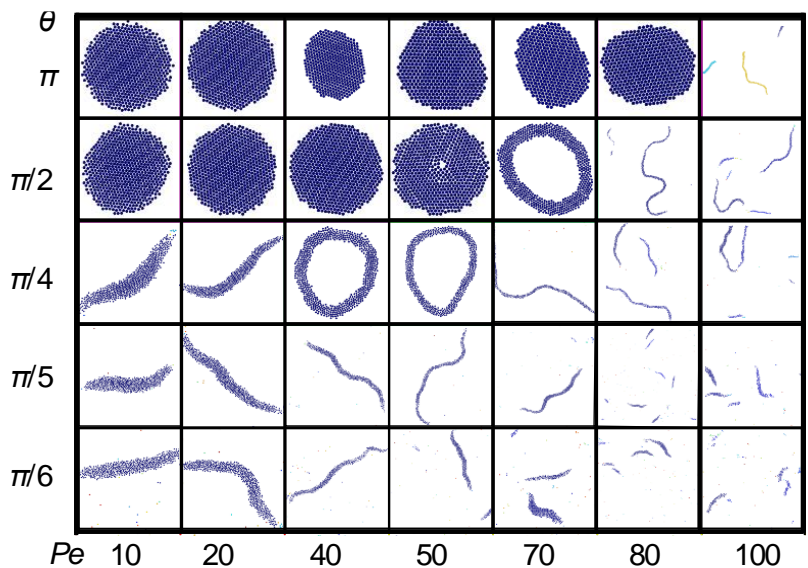

Figure 2: Snapshots of emerging structures for different Peclet numbers Pe, vision angles $\theta$, with alignment-vision ratio $\Omega_a/\Omega_v = 4$ and packing fraction $\phi= 0.00785$. The snapshots are not to scale for better visualization. Adapted from Ref. [7] with permission.

**Current and Future Challenges**
The prediction of the emergent behavior of complex multi-component systems like swarms poses many challenges. This also applies to the search for optimized targeted behavior of microrobotic systems.
***Noise:*** Biological and microrobotic systems are exposed to many kinds of noise: Thermal noise for nano- and microscopic systems, active noise due to the operation of the propulsion machinery, noise in the sensing system, as well as environmental noise affecting sensing, and finally noise due to the variation of properties and behaviors of individuals. Noise is an ubiquitous challenge in pursuit, swarming, and sensing of predators, but can also be exploited by prey for escape.
***Directional Sensing:*** Directional sensing has been mainly studied so far in terms of an idealized visual perception. However, visual perception requires the recognition and processing of optical signals. This puts a lower limit to the size of the organisms, which is a few millimeter for insects and fish (zebrafish larvae). For smaller organisms, chemotactic sensing is probably the dominant mechanism. This form of directional sensing is currently under investigation, but needs more attention.

***Memory and Adaptation:*** Current models typically focus on sensing of the location and orientation of targets or neighboring agents. This certainly makes sense in the spirit of minimal modelling. However, the sensing and processing capabilities of agents should be extended to include some (short-term) memory, which would allow the determination of direction and magnitude of the velocity of other agents. In this way, predictions can be made about locations of these agents in the future, and the own motion can be adapted accordingly. Furthermore, speed adaptation in additional to adaptation of direction of motion can strongly enhance the spectrum of reactive behaviors.

***Cognitive Abilities:*** There are many important and interesting questions concerning cognition, biological and artificial intelligence, the similarities and differences of information processing in the brain function and in a computer, the existence of a free will, etc. For example, the difficulty to distinguish between biological (brain) and artificial (computer) intelligence is emphasized by the Turing test, ChatGPT, etc.

The concept of cognition was initially coined for the human brain. However, it is under intense debate, and has been extended to aneural systems, including bacteria, plants, even to all living organisms [10], and similarly to autonomous machines and (micro)robots. Thus, at the current stage of research, cognition can be considered as a deterministic process (a "strategy"), which transforms the input signal into action – as it can be performed, e.g., by a computer in an autonomous car, or by the biochemical signalling processes in a neutrophil chasing a bacterium. It will be important (i) to elucidate how far this deterministic (minimal) cognition is sufficient to understand and predict the behavior of simple animals, and (ii) to extend the model to incorporate cognitive processes of increasing complexity.

***Hydrodynamics and Finite Reynolds Numbers:*** Many biological agents live in an aqueous environment, like sperm, neutrophils, bacteria, and marine microorganisms. Furthermore, fish schools provide the most impressive examples of swarming behavior. Thus, the interplay of cognition and steering with hydrodynamic propulsion and interactions needs to be understood. An example is the hydrodynamic starvation of first-feeding larval fishes, as low-Reynolds-number hydrodynamics limits their feeding performance. Furthermore, with increasing agent size, inertia effects become important, and hydrodynamic studies should be extended to finite Reynolds numbers. Finally, can a predator invent a strategy to evade the propulsion-induced flow fields of the target which hinder it to reach the prey?

***Modelling of Real Biological Swarms:*** Animal swarms of some species show similar behaviors, while others are very different. Thus, generic models (as the ones described above) have to be complemented by more specific models for selected species.

**Advances in Science and Technology to Meet Challenges**

Advanced simulation codes on the new generation of exascale computers will facilitate the numerical study of several of the challenges mentioned above. Advances in three-dimensional imaging and tracking will provide more detailed information about animal swarms. The development of multifunctional microrobots can pave the way to swarm robotics.

**Concluding Remarks**

The swarming of many identical self-propelled and self-steering agents – without a leader or an external governing force – is a fascinating phenomenon. Generic, minimalistic models derived from the behavior of animal swarms have been developed and analyzed, and highlight the physical principles driving swarm formation. Systems of interacting autonomous cognitive units will provide an important technological route to the organization of complex active systems without central control – also known as distributed computing.

**Acknowledgements**

Many stimulating discussions and collaborations with Marielle Gaßner, Segun Goh, Priyanka Iyer, Rajendra Singh Negi, and Roland G. Winkler are gratefully acknowledged.

# 15 -- Guiding synthetic active particle self-organization with local cues.

Frank Cichos, Molecular Nanophotonics, Leipzig University, 04013 Leipzig Germany

**Status**

Synthetic mobile active microparticles enable the conversion of local energy of different forms into directed motion in a microscopic world. They have opened an exciting new avenue in experimental active matter research, where the active components can be precisely designed and studied to explore non-equilibrium phenomena that find equivalents biological processes, but without the enormous complexity of the biological world. As such, these systems have provided new insights into the fundamental physics of active matter, involving new types of dynamical phase transitions [1], flocking [2], traveling density waves [3], and many other collective effects.

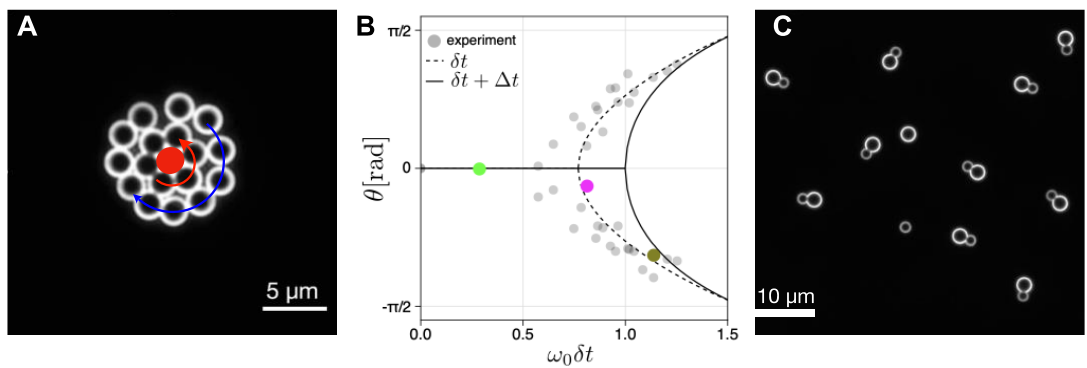

*Figure 8. Emergent dynamics and function by feedback control:* **A** Feedback controlled active particles that show spontaneous rotation around a central particle due to a perception reaction delay *[8]*. The transition to the rotational phase is governed by a pitchfork bifurcation, **B**, combining particle activity ($\omega_0$) and time delay $\delta t$. **C** Colloidal reservoir computer using single active particle rotators to provide the non-linear dynamics for physical reservoir computing *[9]*.

However, the interaction of biological agents, e.g. bacteria, their ability to adapt and learn, goes beyond the simple physical interactions that determine the dynamics of synthetic active particles, but rather involves the exchange, processing and response to the information gathered. In contrast, perception, information exchange and feedback are missing in synthetic active particles. To overcome this deficiency, feedback control of single and many particles was employed, using a camera to perceive the active particle state, a computer to process the information about their state, and the active particles to perform the actions in the physical world. Using such techniques, active Brownian motion rectification [4], bound states of active particles [5], quorum sensing and swirling [6] have been demonstrated because of implemented feedback rules. Algorithms of reinforcement learning have been coupled to real world swimmers to solve simple navigation problems [7]. Feedback control has been implemented to introduce perception reaction time-delays to synthetic active particles, introducing inertia-like dynamics to overdamped systems [8]. The inherent, non-reciprocal, time-delayed interaction also enables the use of active colloidal particles for physical reservoir computing and thus facilitates information processing [9].

Experimental feedback control therefore compensates for the lack of complexity of synthetic active particles, but our understanding of self-propulsion and interactions may now open routes to introduce complexity. This requires the integration of active particles in multi-component systems, as a feedback process, for example, involves several components. Such processes become possible when active particles drive passive soft matter out of equilibrium and this soft matter acts back on the self-organization of the active particles. Therefore, a primary experimental challenge is to increase the complexity of synthetic active particle systems and enable them to self-organize into hierarchical structures [110] based on well-controlled local cues provided by active particles and their soft matter environment.

**Current and Future Challenges**

**Complexity Challenge**
Biological systems, beside replication, encompass a vast array of predominantly passive but functional components with specific interactions, autocatalytic chemical reactions, alongside active components that couple to those for structure formation, transport, and function. The challenge is now to purposefully design multicomponent systems involving synthetic active particles, physical/chemical sources/sinks (Fig. 2A), passive particles (cf. Fig. 2B), droplets (cf. Fig. 2C), functional molecules, chemical reactions, or other species to increase the complexity of the system. Active and passive systems can be equipped with molecular components for specific interactions with controlled binding energy to tune coupling, allowing phase changes or mechanical deformations. A bottom-up design that can be controlled in all its components and understood will thereby allow to understand the emerging dynamic systems and add a new twist to the comparison of biological and synthetic active matter.

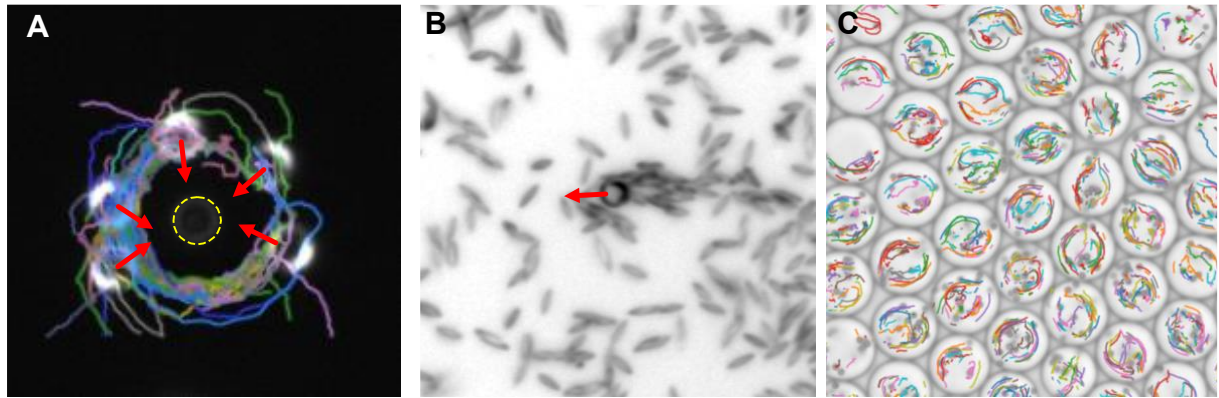

*Figure 9. **A** Radial thermo-osmotic flows and temperature fields create attractive, repulsive fields and hydrodynamic polarization for self-thermophoretic Janus particles (Lisa Rohde, to be published). **B** Induced nematic ordering of ellipsoidal polystyrene particles combining thermally induced depletion forces in a polyethylene glycol/water mixture behind a self-thermophoretic Janus particle (Lisa Rohde, to be published). **C** Optically heated iron oxide doped polymer particles in water droplets induce Marangoni flows (trajectory overlay) to self-organize droplets in large arrays synchronizing their flow fields (Akshay Kallikkunnath, to be published).*

**Cues for Self-Organization Challenge**
An important challenge on the way for more complexity is to identify suitable local cues for the self-organization that set also define characteristic length scales for a hierarchical ordering of active and passive components and of into structures with new emergent properties. Synthetic active particles react to many influences that can provide their self-organization. Inhomogeneous hydrodynamic flows, composition gradients, temperature fields and chemical fields are just some of them, which are and need to be further explored. Here, synthetic active matter research may converge with new approaches of thermoplasmonics, where a variety of temperature-induced effects, such as thermophoretic drifts, thermo-osmotic flows, thermo-viscous flows, thermally induced composition fields or thermo-electric fields have recently gained considerable attention and allow precise control of particles, bacteria or even molecules. Local cues from optically pumped plasmonic structures that generate a local temperature induced osmotic and phoretic processes can order synthetic active and passive matter by dynamic processes, as shown in Fig. 2A-C for increasing complexity. These examples can serve as building blocks for use in larger ensembles. Since these thermally induced processes have analogies to other physical or chemical fields, they may serve as a well-controlled model system.

**Soft Active Particle Challenge**
While most of the current synthetic active systems are colloidal particles and comprise hard sphere or DLVO interactions, deformable synthetic active systems are rare. Chemically self-propelled emulsion droplets, deformable vesicles filled with synthetic active particles that exhibit large fluctuations in shape, or thermally driven droplets (Fig. 2 C) are examples that can provide various cues for self-organization on their own but can also react to mechanical cues in the environment, e.g. by deforming and jamming, for example. Combining the properties of emulsions or micelles with activity provided by synthetic components is therefore a very exciting challenge that can deliver large complexity for synthetic active matter systems.

**Advances in Science and Technology to Meet Challenges**
Creating ensembles of active particle, diverse in shape, composition, propulsion mechanisms together with other passive colloidal or molecular components for the study of complex self-organization need multiple advances.
**Advancement in Experimental Tools** The greater complexity of the systems makes it necessary to go beyond simple optical imaging to gain access to all components. Multimodal imaging joining the feedback control techniques with fluorescence, phase contrast, chemical imaging, and even three-dimensional resolution, e.g. through optical holography or confocal sections, or advanced computational imaging is required.
**Advancements in Data Analysis** The multicomponent systems also require new approaches for data analysis, as tracking individual particles in dense multicomponent systems may not be possible and other approaches are needed to characterize the dynamic state of the system. Here, novel machine learning tools that segment images and reduce the dimensionality of the observed sample state will play an important role in data analysis.
**Materials Advances** Coupling synthetic active matter to passive and molecular components needs new designs, that involve new technologies, like 3D printing, colloidal synthesis, DNA nanotechnology for binding energy engineering and others to provide precise control of the systems.

**Concluding Remarks**
Synthetic motile active particles are often recognized as simple agents that mimic a basic functionality of bacteria and deliver a glimpse of the physics of non-equilibrium collective effects that are omnipresent in living systems. Yet, synthetic active particles are sources of mechanical activity that can power soft matter systems to self-organize into systems of increasing complexity based on local physical and chemical cues that need to be explored. The increased complexity of synthetic active matter is required to create emergent functionalities involving feedback and information exchange, which are currently only introduced by experimental feedback control, but play a vital role for their living counterparts.

**Acknowledgements**
I would like to thank Lisa Rohde, Akshay Kallikkunnath, Gordei Anchutkin, and Xiangzun Wang for contributing unpublished results and many discussions to the topic of the publication. We acknowledge funding by German Research Foundation (Deutsche Forschungsgemeinschaft, DFG) through project no 432421051 and by the Center for Scalable Data Analytics and Artificial Intelligence (Scads.AI) Dresden/Leipzig.

# 16 -- Microrobotic Swarms Trained by Reinforcement Learning

Veit-Lorenz Heuthe and Clemens Bechinger
Physics Department, University of Konstanz, Germany

**Status**

In many animal species, individuals cooperate and use collective strategies that allow the group to overcome the limitations set by individual efforts. Dolphins, for example, hunt in groups that can herd prey more effectively, termites build intricate nests to regulate their climate, and ants in a colony organize to transport heavy objects together. Inspired by such natural examples of collective behaviour, roboticists are designing platforms in which many copies of the same, unspecific robots work together to solve complex tasks. Opposed to single complex robotic units, the advantages of such swarm robotic systems (SRS) are their flexibility, scalability and robustness which arise from decentralized control and the low degree of robotic specialization and complexity: A swarm of robots can take on many different spatial configurations and can therefore be adapted to a different task by simply reprogramming the units. To change the size of the system, simply add more robots. And if a unit fails, it can be replaced by another one without affecting the function of the whole group. To date, various types of SRS have been successfully realized with robots at the centimetre scale and above, that are for example able to spontaneously organize into different patterns [1] or construct larger structures [2]. At the microscopic scale, SRSs hold great promise for minimally invasive medicine, water purification, and directed assembly of small structures [3]. Recently, it has been shown that magnetic, electric and acoustic fields can be used to activate colloids as microrobot swarms, which can organize themselves into different patterns or transport larger cargo colloids [4, 5].

**Current and Future Challenges**

Most current microscopic SRS are actuated by fields that extend over the entire swarm. In these systems, collective functions arise from the dominant local interactions (e.g. magnetic dipole interactions), in contrast to macroscopic SRS, where the behaviour of individual units can be arbitrarily programmed. To explore the full potential of microscopic SRS, however, individual control and dedicated collective task solving strategies are required. Aside from the challenge of miniaturizing energy storage, sensors, and electronics, the biggest challenge for the development of microscopic SRS is to develop control schemes at the level of individual robots that produce the desired behaviour at the level of the swarm, and that can handle the high levels of noise inherent in the microscopic length scale. Recently, automatic design methods and specifically Multi-Agent Reinforcement Learning (MARL) algorithms have gained much attention as a powerful platform for addressing the issue of SRS control [6]. Within the framework of MARL, control schemes on the scale of individual agents (i.e. robots) can be derived based on user-defined collective goals and by employing unsupervised learning algorithms aiming to optimize each robot's (agent) behaviour with respect to a performance-based reward. This approach effectively translates the challenge of programming individual control schemes for each robot inside a swarm into finding an appropriate reward function. There are various common approaches with different strengths: Simply rewarding each agent based on the performance of the group leads to high variances in the reward signal that slow down training and can lead to suboptimal results because the agents cannot discern rewards due to their own or the other agents' effort. On the other hand, manually designing reward functions for individual robots requires a good understanding of the specific task and environment, the specific choice of the reward function can bias the trained strategy and the reward function needs to be redesigned for different tasks. In summary, effective rewarding schemes need to assign credit to each individual robot without requiring an assumption about the ideal strategy beforehand. A possible solution to this problem is the concept of counterfactual rewards: In this rewarding scheme, the contribution of each agent is determined by the difference in the swarm performance with and without the participation of the agent [7]. In practical terms, this comparison of real and counterfactual scenarios can be achieved by running real experiments and numerical simulations in which individual agents can be removed or made passive simultaneously. The use of counterfactual rewards also reduces the variance in the rewards for each single agent compared to that of the entire swarm. So far, several MARL algorithms were constructed on this basis but experimental implementations in SRS are sparse. This is particularly true for microscopic robots which hold great potential in view of the above-mentioned applications, but which typically suffer from a less deterministic motion than their macroscopic counterparts.

**Advances in Science and Technology to Meet Challenges**
To enable a microscopic SRS to perform a specific task, robots must sense their environment and must be able to adjust their motion accordingly. While corresponding sensors and self-driving units can be implemented on board in case of larger robots (e.g. Kilobots), such requirements are currently difficult to meet at microscopic scales. As an intermediate step towards this ambitious goals, one can move the sensing and steering capabilities from the microrobot to external units which provide the robots with the information regarding their surrounding but also deliver the mechanism for their motion. As an example, one can use optical feedback controls, where an online particle tracking algorithm enables the permanently updated information regarding the position and orientation of peers. This information can be coupled to a scanning laser beam which illuminates each particle such to yield a specific direction and magnitude of the robotic velocity. This approach has been already demonstrated to in connection with deterministic controlled interaction rules applied to micron-sized active colloidal particles [8]. When combining this approach with the above-described MARL methods, we expect the training of micro robotic groups towards collective tasks to be within reach [9, 10].

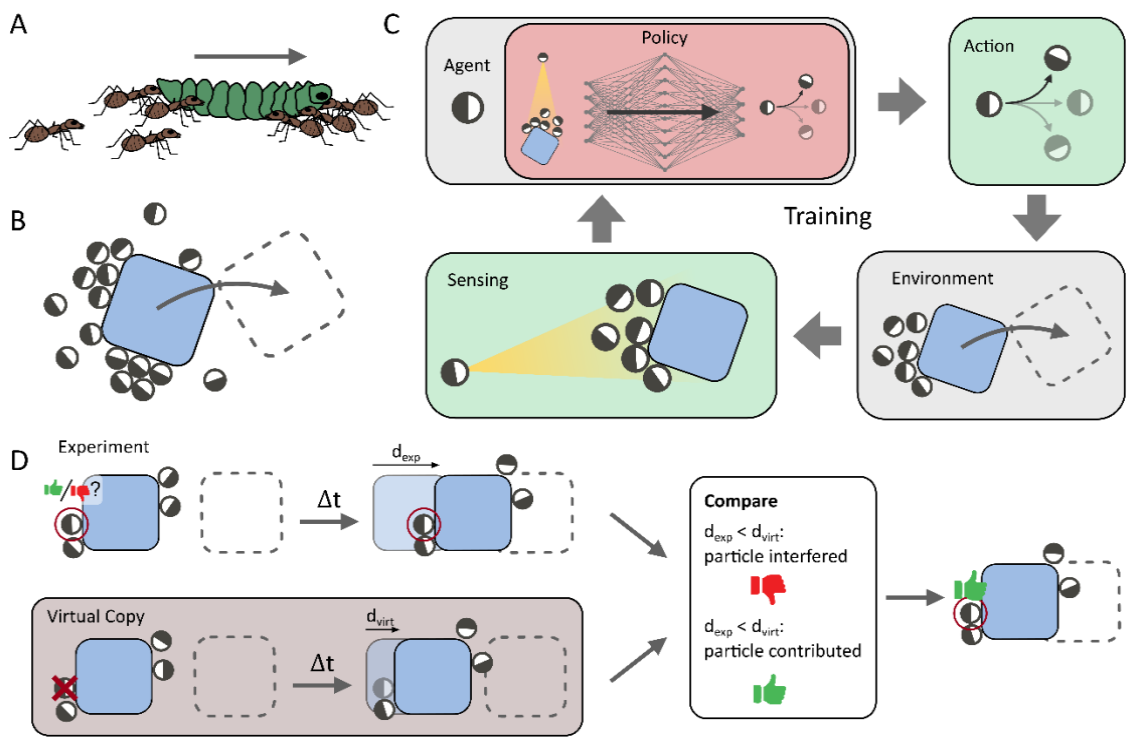

*(A) Animals use collective strategies to overcome the limitations of the individual, e.g. to transport large objects. (B) Similarly, microscopic robot swarms could solve collective challenges. (C) In Multi-Agent Reinforcement Learning (MARL) an optimal control scheme for solving a collective task is automatically obtained by training the policies of the individual agents (microswimmers). The agents learn by exploring their environment in successive*

*sensing and acting steps. **(D)** For training in MARL, the individual agents need to be rewarded. For counterfactual rewards, each agent is once removed virtually; if this decreases the swarm's performance, the removed agent did contribute positively and thus gets a high reward and vice versa.*

## Concluding Remarks

Advances in active matter research have brought the development of swarms of thousands of microrobots that coordinate their motion and solve complex tasks within reach. This raises the question of how to devise control frameworks for each individual microrobot that enable the group to do something useful. Multi-agent reinforcement learning is a promising approach to this challenge, but needs to be combined with rewarding schemes that allow it to play to its strengths. By implementing counterfactual rewards together with MARL, it is possible to achieve systems that are easily adaptable to different tasks, scalable to many thousands of microrobots, and robust against the high levels of noise that are inherently present at the micro-scale.

## Acknowledgements

*This work was supported by the DFG Center of Excellence 2117 'Centre for the Advanced Study of Collective Behaviour' (ID: 422037984)*

# 17 -- Collective Problem Solving by Swarms

Amos Korman[1] and Ofer Feinerman[2],
[1]University of Haifa, Department of Computer Science, Haifa, Israel, and the French National Center for Scientific Research (CNRS), UMI FILOFOCS, Israel
[2]Department of Physics of Complex Systems, Weizmann Institute of Science

**Status**

Cognition describes the process by which individual animals acquire, process, and act upon internal and external information. The term has since been expanded to encompass the collective cognition displayed by animal groups (Couzin, 2009). This expansion is not merely semantic since the collective dynamics in cognitive groups often mirror the neuronal dynamics in animal brains that confront comparable challenges.

Biological models of swarm cognition tend to consider swarm members as very simple entities: Ants are modeled as magnetic spins (Feinerman et al., 2018) and fish as on-off light sensors (Berdahl et al., 2013). As such, there have been attempts to draw similarities between emergent group dynamics and computation in neuronal circuits. Within such comparisons, animal groups with simple members that are connected via ad-hoc, temporary, anonymous, mobile, and noisy networks could be no more than a pale shadow of the corresponding designated, stable, identified, tunable, and embedded neuronal circuits (Boczkowski et al., 2018). This might lead us to expect swarm cognition to be "less than the sum of its parts" or, even worse, less than one part.

That being said, the basis for the simple-swarm-member assumption is not completely clear. Ants, fish, and birds all display significant cognitive capabilities as individuals. Employing these capabilities in a swarm setting might help the group increase its collective performance. This possibility is backed by theoretical studies which provide examples in which groups can cooperate optimally but only if each agent has a large memory that increases with group size (Feinerman & Korman, 2017). In swarms of eusocial animals, and to a lesser extent other swarms, group performance is important for survival. This suggests that individuals in the swarm could sometimes employ their own cognitive skills in a way that maximizes collective cognition (Sasaki & Pratt, 2012), such that it scales efficiently with group size, and allows for swarms that are "more than the sum of their parts" (Berdahl et al., 2013).

Cognitive puzzles provide an efficient means for measuring cognition. Puzzles capture the problem solving, goal oriented dynamics that distinguish living from physical systems. Indeed, cognitive puzzles have been repeatedly employed to characterize and quantify single animal cognition. Groups of animals could also be challenged by puzzles: Argentine ants use mass recruitment trails to solve the famous Towers of Hanoi puzzle (Reid et al., 2011) while crazy ants employ cooperative transport to outperform conventional solutions of the Ant-in-a-Labyrinth puzzle (Gelblum et al., 2020). Thus, puzzles could provide us with empirical measures of comparing cognition across scales.

**Current and Future Challenges**

The overarching challenge is to elucidate the connections between individual and group cognition and develop a theory which quantitatively captures the correct scaling relations. A main goal towards this is to identify scalable puzzles in which groups apply qualitatively higher cognitive processes to impressively outperform individuals.

There are inherent conceptual difficulties in comparing problem solving skills across organizational scales. A neuron, for example, acts in a world of neurotransmitter clouds, ion currents, and action potentials and works, for instance, to maximize information flow rates. These are in no way comparable to the environment that a brain composed of such neurons must operate in. Similarly, it is not clear how one could compare emergent swarm intelligence to individual animal cognition. In fact, a common definition views emergence as a situation where a new vocabulary is required to describe higher hierarchies (Johnson, 2002). Thus, the mere definition of emergent collective cognition might defy the formulation of scaling laws.

Further challenges arise when considering the simplified models used to describe animals in swarms. Roughly speaking, we do not know which of the following holds: Either the models are oversimplified and reflect our own limitations rather than biological reality, or that models do capture reality and hence, when part of a swarm, animals do not utilize their cognitive capacities to the full.

If it is indeed the case that animals in swarms behave in a simplistic manner then an additional challenge would be to understand the reasons for this. Solutions may come from several directions: one possible explanation is that coordination is difficult to achieve, since, e.g., communication between freely moving swarm animals is highly noisy and unpredictable (Boczkowski et al., 2018). This not only imposes limitations on coordination but might also make complex computations on the individual scale redundant, as their results cannot be disseminated. A second explanation might be that to work in concert animals may attempt to be predictable to one another, and that this predictability requires simplicity.

**Advances in Science and Technology to Meet Challenges**
Unlike the neurons-versus-brain example presented above, animals, whether alone or as a group, often occupy the same environment, face similar challenges, and perform similar tasks: Birds navigate alone or within a flock, solitary moles search for food alone but naked mole-rats search as a group, ants carry small loads while ant teams cooperatively transport large insects (Figure 1). Moreover, since group behavior evolved after individual cognition, the mental resources and problem solving capabilities required by the group are often present in each individual. One could therefore envision designing scalable puzzles that could be presented to either individuals or groups (Figure 1). Identifying such puzzles would constitute a first, necessary step towards a quantitative study of the scaling properties of collective cognition.

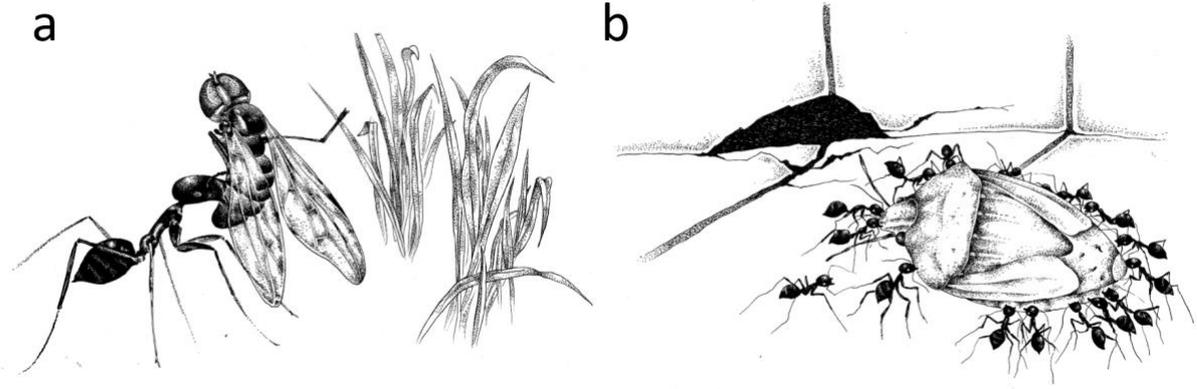

**Figure 1:** Longhorn crazy ants confront natural puzzles. Depending on the size of the load these ants retrieve food either independently (a) or as a group (b). The challenge of maneuvering the load to the nest through tight passages is ecologically relevant in both cases. Such natural challenges can inspire scalable puzzles that could allow for a quantitative study of collective cognition.

What would be the design principles of scalable puzzles? Puzzles should be natural, pressing, and ecologically relevant at all scales. These properties will ensure that the puzzles engage the solver's (individual or group) attention and interest, and that the solver is equipped to solve them at a level that has passed the test of evolution. There is an advantage to puzzles that allow different levels of difficulty: Collective cognition may be quantified by the most difficult puzzle which is solvable per group size. Finally, while measurements of collective performances may be teaching, the route towards understanding involves drawing connections between the organizational scales. This sets the experimental requirement to track all individuals within a group, quantify their behavioral states, and eavesdrop on their communication during the cognitive process. Recent technological advances are making such detailed measurements increasingly feasible.

A final design principle concerns theory and puzzle design. Specifically, the natural language for studying the inherent computational challenges faced by the group is that of theoretical computer science. Although this community has traditionally focused on engineering applications, recently, attempts have been made to study bio-inspired models from a distributed computing perspective (Zhang et al., 2022). One possible application of such studies is to quantify the difficulty of a puzzle by evaluating the resources required to solve it efficiently. While this may necessitate assumptions regarding coordination, a benchmark could be the case where communication is centralized and unlimited. Another application is to identify those puzzles that when solved by a group, require sophistication on the individual level. If animal groups efficiently solve these puzzles this would defy the common assumption of simple individuals and could allow for group performances that are particularly impressive.

**Concluding Remarks**
To date, most swarm intelligence models treat animals as indifferent physical particles. Collective cognition arises as the interaction network and collective dynamics in the group form computational circuits reminiscent of neuronal circuits, albeit with a significantly less structured communication scheme. Since these approximated neuronal circuits are, in fact, present in the brain of every animal within the swarm then it would appear that when compared to individuals, groups are expected to display decreased cognitive abilities.

We suggest a collective problem solving paradigm as a means of empirically quantifying group cognition and searching for examples where it reaches its most impressive limits. It may very well be that inherent communication constraints force animals to simplify their decisions and communication in order to be able to achieve coordination. On the other hand, we believe that the most impressive examples of group problem solving would rely on individuals that employ their cognitive capabilities to the full. Identifying those puzzles that allow for the quantification of computation on both the individual and group levels remains a main challenge.

## Acknowledgements
This work has received funding from the European Research Council (ERC) under the European Union's Horizon 2020 research and innovation program (grant agreement No. 770964). OF is the incumbent of the H. J. Leir Professorial chair. We thank Sarit Barnard for the illustration in figure 1.

# 18 - Renormalization group crossovers in living systems: when finite size matters

Andrea Cavagna[1,2] and Irene Giardina[2,1],
[1] Istituto Sistemi Complessi (ISC-CNR), Rome, Italy
[2] Dipartimento di Fisica, Sapienza Università di Roma & INFN, Unità di Roma 1, Rome, Italy

**Status**

The collective behaviour of active living matter seems to befit some of the conceptual structures of statistical physics, but to pursue this idea the great diversity of biology needs to be drastically simplified. The Renormalization Group (RG) – with its classification of relevant vs irrelevant variables – is optimally suited to this task and the first steps taken along this path make us optimistic [1–4]. It is therefore timely to speak a word of caution about future RG studies of living systems. The fact that the limit of the RG flow is a certain attractive fixed-point means that the infinite length-scale physics of the system is ruled by that fixed-point, as the process of iterating the RG transformation corresponds to observing the system at larger and larger length-scales. Biological aggregates, however, are quintessential finite-size systems, in which case it becomes crucial to take into account the potential role of RG crossovers.

**Current and Future Challenges**

Consider a microscopic system close to some RG fixed-point, A, that is unstable with respect to one parameter, let us call it $x$: the RG flow along $x$ leads from A to some attractive fixed-point B. Because $x$ is a relevant variable growing away from A, the fact that the system is in the vicinity of A means that the physical value of $x$ is small. Which one between A and B will rule the physics of the system? It depends on two things: the physical value of $x$ and the system's size $L$. The value of $x$ determines a crossover length-scale, $R_c(x)$, below which A still rules the physics of the system and beyond which B does. This scale, $R_c(x)$ is larger the smaller is $x$, and it goes to infinity when $x = 0$: if we kill the unstable parameter, A becomes an attractive fixed point. In the thermodynamic limit we will necessarily cross over to the regime $L \gg R_c$ and the stable fixed-point B will dominate. On the contrary, if the system has finite size and $L \ll R_c$, the system is dominated by the unstable fixed-point, A. We will illustrate this crossover in three cases of some biophysical interest.

*First crossover: from inactive to active.* The Vicsek model [5] is a Heisenberg ferromagnet on the run: each 'pointer' aligns to its neighbours and moves its position by following its own direction. One would expect that activity changes the universality class of Vicsek with respect to Heisenberg. However, if particles move with very, very small speed $v_0$, one expects to recover the Heisenberg model. This is where the RG crossover comes into play. Within the RG analysis of the incompressible Vicsek class at criticality [2], the activity coupling constant, $\alpha$, is a relevant variable: the inactive fixed-point ($\alpha = 0$) is unstable with respect to $\alpha$, so the RG flow goes to a stable active fixed-point. The crossover is ruled by a length-scale, $R_c(\alpha)$, which is infinite for $\alpha = 0$ and decreases for increasing activity [6]. For very slow particles, i.e. for $\alpha \ll 1$, we have a large crossover scale, so that finite-size groups may have $L \ll R_c(\alpha)$, and the universality class of these systems will be that of the inactive fixed-point. On the contrary, for large activity the crossover scale shrinks, so that $L \gg R_c(\alpha)$ and the new active fixed-point dominates (see Fig.1a). Natural swarms of midges are on the active side of this crossover [4]: the relaxation time of the velocities is of the same order as the rearrangement time of the interaction network, indicating that these systems are close to the active fixed-point. This is indeed confirmed by the agreement between the RG value of the active critical exponent and the experimental one [4]. For flocks of birds, on the other hand, it is still unclear whether activity changes the universality class; on the one hand, within a strongly polarized system the time to rearrange the interaction network is very large (because the speed in the centre of mass reference frame is small); on the other hand, the continuous broken symmetry implies that the collective relaxation time diverges, hence in the long time limit activity could matter. The jury is still out on that.

*Second crossover: from underdamped to overdamped.* Experiments on bird flocks and midge swarms indicate that dynamic relaxation is underdamped, suggesting the presence of behavioural inertia: the alignment force produced by the neighbours, instead of acting directly on a particle's velocity, as in the Vicsek model, must act on the generator of rotations of the velocity, or spin [8]. Because the inter-individual interaction is rotationally symmetric, the total spin is conserved, but interaction with the environment may dissipate the spin with some friction, $\eta$. Friction grows along the RG flow, asymptotically reaching the overdamped limit (i.e. the Vicsek model), which therefore rules in the thermodynamic limit [4] (see Fig. 1b). Yet again, the flow is regulated by a crossover length-scale $R_c(\eta)$, and as long as $L \ll R_c(\eta)$ the unstable underdamped fixed-point will rule the dynamics. This is exactly what happens for bird flocks and insect swarms [4, 7, 9]: friction is small, so that $R_c(\eta)$ is large enough for these finite-size systems to belong to the underdamped universality class (Fig. 1b). In flocks this fact is at the basis of traveling waves uncoupled to density fluctuations [9], while for natural swarms this conclusion is confirmed by the fact that the dynamical critical exponent of the (unstable) underdamped fixed-point is in excellent agreement with the experimental value [4]. Hence, when it comes to underdamped vs overdamped dynamics, natural swarms and flocks are on the unstable side of the RG crossover.

*Third crossover: from homogeneous to heterogeneous.* It is pointless to apply the RG to systems undergoing a first-order phase transition, and yet we know that the ordering transition in Vicsek active matter is indeed first-order, due to the formation of density heterogeneities caused by the feedback between activity and alignment. Instead, the incompressible case, where density is constant by constraint, has a second-order transition with a well-defined RG fixed-point [2]. The formation of density fluctuations corresponds to an RG relevant direction, with respect to which the incompressible fixed-point is unstable. This crossover has been studied in [10], within a theory in which the number of particles is free to fluctuate [11], so that births and deaths quickly relax local density fluctuations, partially suppressing them. Within an RG analysis this balance is tuned by a parameter $\zeta$ that grows with density fluctuations; for $\zeta = 0$ the system is in the incompressible universality class, but $\zeta$ is a relevant parameter, so that this fixed-point is unstable with respect to $\zeta$. Once again, a crossover length-scale exists, $R_c(\zeta)$, such that finite-size systems with $L \ll R_c(\zeta)$ belong to the universality class of the homogeneous incompressible fixed-point, while within systems with $R_c(\zeta) \ll L$, density heterogeneities take over. To understand on what side of the crossover we are, we can look for hallmarks of second-order physics, as scaling laws and scale-free correlations, and study how strong are density fluctuations in the system at hand. By doing this, solid experimental evidence was gathered that both natural swarms of midges [4] and flocks of birds [9] belong to the neighbourhood of the homogeneous fixed point.

**Concluding Remarks**

The program to classify active matter systems in terms of RG universality classes is moving its first steps. Whenever a new RG fixed-point is discovered, we should ask: is there a crossover associated to it? And, if yes, on what side of the crossover are we? We have seen that the situation in natural swarms and flocks is complex: both swarms and flocks are on the unstable side regarding inertial dynamics and density homogeneity; on the other hand, swarms are on the stable side of the activity crossover, while it is still unclear what side of this crossover flocks inhabit.

**Acknowledgements**

We thank L. Di Carlo, T. Grigera, G. Pisegna, and M.Scandolo, for having accompanied us in the discovery of the relevance of RG crossovers in biological systems. This work was supported by ERC grant RG.BIO (n. 785932) and by PRIN2020 Grant n. 2020PFCXPE.

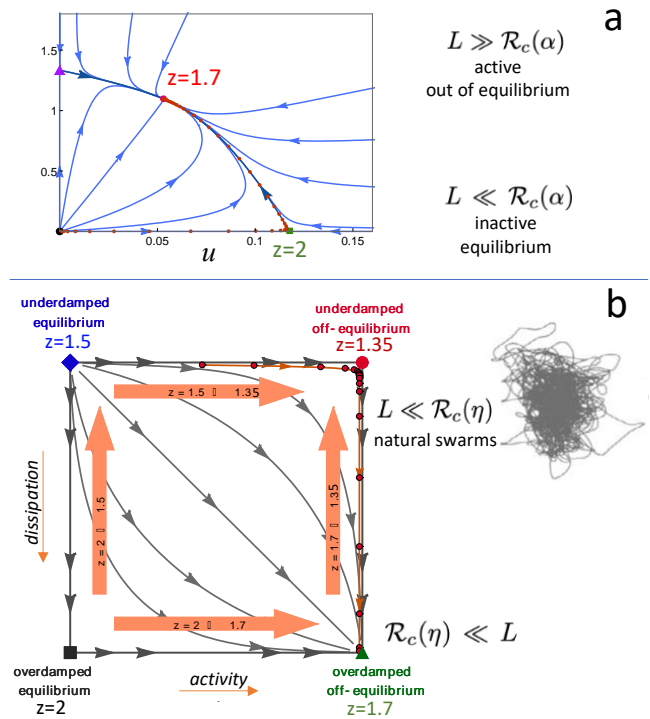

Figure 1. a) Inactive to active crossover: RG flow for the incompressible Vicsek model in d = 3; α is the activity effective constant and u is the interaction effective constant (see [2, 6]). b) Schematic RG diagram of the ISM field theory of [4].

# 19 -- Gene expression and phenotypic heterogeneity during bacterial swarming

Hannah Jeckel[1] and Knut Drescher[2]

[1] Department of Biology and Biological Engineering, California Institute of Technology, 91125 Pasadena, USA

[2] Biozentrum, University of Basel, 4056 Basel, Switzerland

**Status**

Motile active matter systems range in scale from macroscopic flocks of animals to microscopic communities of self-propelled eukaryotic cells, bacterial cells, and abiotic motile Janus particles. Non-living active matter systems are typically composed of a fixed number of particles, which may change their behavior as a direct consequence of their local environment. In contrast, living systems typically contain a varying number of agents whose morphology and behavior, referred to as their phenotype, is not only influenced by their current local environment, but also influenced by the previous environments that each agent has experienced during different times of its organismal development. The information of the current and previous environments that an agent has experienced during its development is transferred into the current phenotype of the agent through gene regulatory mechanisms. As a result of this complexity of motile active matter systems comprising living agents, there can be several distinct phenotypes of genetically identical agents present at the same time in the same local environment, leading to a rich diversity of interactions between phenotypic subpopulations that may give rise to emergent behaviors of the active matter system.

Even for the biologically simplest motile active matter systems, such as bacterial swarms, it is unclear how gene expression causes the formation of different phenotypic subpopulations, and how the interactions of these subpopulations influence emergent behaviors, such as swarm expansion and susceptibility to antibiotics. This section discusses bacterial swarming as a model system to study spatiotemporal phenotypic heterogeneity in motile active matter. Furthermore, this section highlights recent advances as well as future challenges for characterizing the phenotypic subpopulations that arise in bacterial swarm communities and how subpopulation behaviors and interactions at the microscopic scale connect to the macroscale collective dynamics.

**Current and Future Challenges**

During bacterial swarming, a densely packed community of self-propelled bacteria rapidly expands across macroscopic distances on a surface (Fig. 1). While this community expands, the number of cells in the community grows, so that the community expansion is driven by motility, growth, and the underlying metabolism that the cells perform to fuel growth and motility [1], [2]. Using microscopy-based observations, it has been comprehensively documented that the cellular behavior and cellular morphology varies drastically between different temporal stages of the swarm expansion process, between different spatial regions within the swarm, and even within the same spatiotemporal location in the swarm [2], [3]: In some regions cells form highly motile rafts, in some regions cells primarily move as isolated cells, or form non-motile cell clusters or cell chains, and in other regions these different cellular phenotypes co-occur (Fig. 2). A major research challenge is to understand how the collective dynamics of the swarm emerge from the microscopic activities of the phenotypically different bacterial cells.

On short time scales, i.e. time scales in which cellular phenotypes do not change and on which the swarm community does not significantly expand, mechanical interactions between the cells are sufficient to qualitatively and quantitatively describe the collective dynamics of the bacterial swarm [2].

On longer time scales, i.e. time scales on which the cellular phenotypes change, which are also the time scales on which the swarm expands significantly and the cell number of the swarm changes significantly, it is unclear how the microscopic cellular activities generate the emergent collective dynamics of the swarm. To develop an understanding of the system dynamics on these longer time scales, an important challenge is that new experimental technologies are needed: While brightfield microscopy methods have been useful in the past for characterizing cellular motility and morphology, they cannot capture the underlying biological reprogramming of cells that results in changes in cellular phenotypes. Adapting or developing methodologies for comprehensively characterizing the gene expression at the cellular level in bacterial swarms are therefore required. Once these methods are available, another important challenge will be the development of data analysis strategies that incorporate temporal and spatial information into existing statistical frameworks, to utilize these spatiotemporal gene expression datasets for formulating models that incorporate gene expression, changes of cellular phenotypes, as well as cell morphology and cell motility. Such models could potentially establish a scale-bridging understanding of how gene expression, microscopic activities of the bacterial cells, and interactions between the phenotypically different cell subpopulations cause the collective dynamics during bacterial swarm expansion.

**Advances in Science and Technology to Meet Challenges**

To understand how gene expression leads to phenotypic heterogeneity during swarming, and how the microscopic activities of phenotypically heterogeneous cells cause the emergent collective properties of bacterial swarms, experimental data needs to be recorded at multiple layers of information. While it is important to record the motility dynamics and cell morphology of all cells in the swarm (which is typically achieved using brightfield live-cell microscopy), information of the gene expression needs to also be recorded for every cell during the swarming dynamics. Currently, this is only possible using fluorescent protein-based gene expression reporters, which enable live imaging at the single cell level. However, this fluorescent-reporter-based approach is limited by the time-consuming requirement of having to genetically engineer a separate bacterial strain for each gene that needs to be read out. Furthermore, due to the spectral limitations of fluorescent proteins, it is typically not possible to read out more than 3-5 different fluorescent proteins at the same time. Despite these limitations, studies using fluorescent reporters for studying gene expression during bacterial swarming have already yielded important insights into swarm development [4].

To overcome the limitations of fluorescent reporters for measuring gene expression during swarming, techniques for measuring gene expression of all genes simultaneously are promising, such as transcriptome or proteome measurements. A decade ago, these techniques required >10^5 cells for a single measurement, which meant that they only allowed bulk measurements of gene expression. However, spatial resolution can still be achieved when these bulk technologies are combined with spatial sampling [5], and a spatiotemporal characterization of large-scale gene expression patterns during swarming was recently accomplished using robotic sampling at a spatial resolution of ~100 µm [6]. Significant advances of sequencing techniques have furthermore made it possible to measure the expression level of 5-20% of the genes (i.e. hundreds of genes) in a single bacterial cell, for example by using multiplexed fluorescence in-situ hybridization techniques based on imaging [7] or the rapidly improving single-cell RNA-seq techniques [8]. These technological developments for measuring genome-wide gene expression levels at the single-cell level could provide the necessary toolkits to enable the simultaneous spatiotemporal measurements of gene expression, cellular motility dynamics, and collective dynamics during swarm development.

**Concluding Remarks**

Understanding the dynamics of biological motile active matter systems comes with unique challenges associated with the phenotypic heterogeneity of living agents, their ability to change phenotypes over time based on changes in gene expression, and their ability to proliferate. To obtain an understanding such motile active matter systems over time scales on which the cells grow and change their phenotypes, we first need a characterization of the system in terms of gene expression, activity of individual cells in terms of motility and growth, as well as the collective cell activity, to provide the basis for developing models that connect processes between these three length scales. Bacterial communities, such as swarms, are particularly suited for achieving such a scale-spanning understanding of motile active matter, because bacteria are relatively simple organisms (compared to eukaryotes), which can be easily genetically manipulated, with fast replication times, which can be well-controlled in laboratory experiments. Furthermore, the availability of detailed annotations of genomes of model organisms for bacterial swarming, such as *Bacillus subtilis* or *Pseudomonas aeruginosa*, can enable interpretations of which intracellular processes influence the motility and morphology of cells, and ultimately the collective dynamics of the whole swarm.

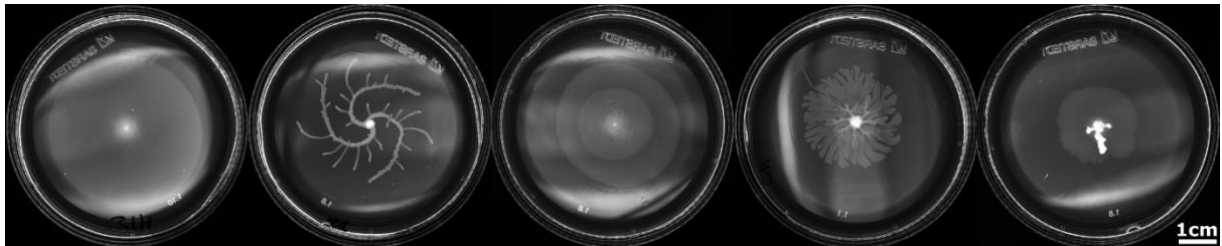

**Fig. 1.** Macroscopic warming patterns created by *B. subtilis* wildtype (leftmost) and different mutants.

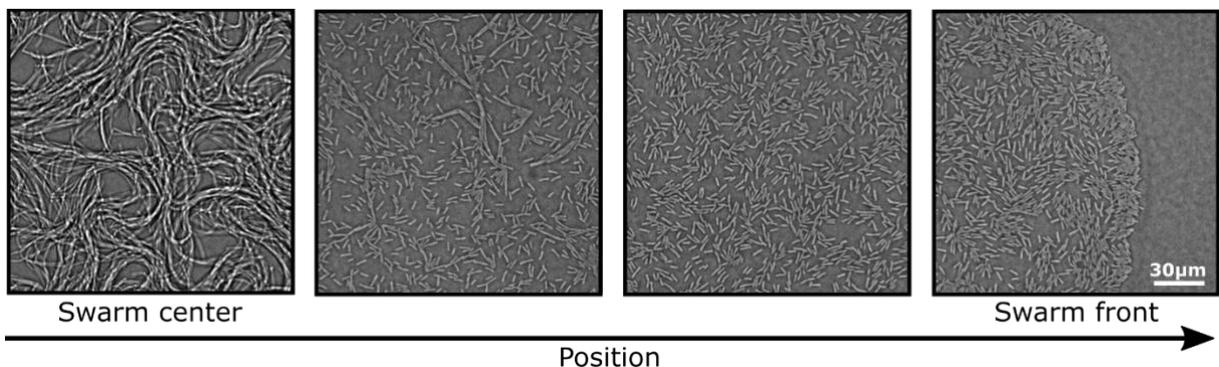

**Fig. 2.** Microscope images acquired at different positions within a *B. subtilis* swarm. This figure shows that the shape of individual cells changes dramatically between the swarm center and the swarm front.

**Acknowledgements**

We are grateful for the following funders for enabling our work on this topic: Swiss National Science Foundation (SNSF Consolidator Grant TMCG-3_213801 to K.D.), Deutsche Forschungsgemeinschaft (DR 982/6-1 to K.D., as part of the Priority Programme SPP 2389).